\documentclass[letterpaper,number,1p]{elsarticle}
\usepackage{times}
\usepackage[T1]{fontenc}
\usepackage{amsmath}
\usepackage{amssymb}
\usepackage{esint}
\usepackage{microtype}
\usepackage{subfig}
\usepackage{booktabs}
\usepackage{amsfonts}
\usepackage{enumitem}
\usepackage{lineno}
\begin{document}
\begin{frontmatter}
\title{\Large Multi-directional Behavior of Granular Materials
              and Its Relation to Incremental Elasto-plasticity}
\author[up]{Matthew~R.~Kuhn\corref{cor1}}
  \ead{kuhn@up.edu}
\author[insa]{Ali~Daouadji}
  \ead{ali.daouadji@insa-lyon.fr}
\cortext[cor1]{Corresponding to:
               Donald P. Shiley School of Engineering,
               University of Portland,
               5000 N. Willamette Blvd.,
               Portland, OR, 97203, USA.
               Email: \texttt{kuhn@up.edu}}
\address[up]{Br. Godfrey Vassallo Prof. of Engrg.,
             Donald P. Shiley School of Engrg., Univ. of Portland,
             5000 N. Willamette Blvd., Portland, OR, USA 97231}
\address[insa]{University of Lyon, INSA-Lyon, GEOMAS, F-69621, France}
\begin{abstract}
The complex incremental behavior of granular materials is
explored 
with multi-directional loading probes.
An advanced discrete element model (DEM) was used to
examine the reversible and irreversible
strains for small loading probes, which follow
an initial monotonic axisymmetric triaxial loading.
The model used
non-convex non-spherical particles and an exact implementation of the
Hertz-like Cattaneo--Mindlin model for the contact interactions.
Orthotropic true-triaxial probes were used in the study (i.e., no
direct shear strain or principal stress rotation),
with small strains increments of $2\times 10^{-6}$. 
The reversible response was linear but exhibited a high degree
of stiffness anisotropy.
The irreversible behavior, however,
departed in several respects from classical elasto-plasticity.
A small amount of irreversible strain and contact slipping
occurred for all directions of
the stress increment (loading, unloading, transverse loading, etc.),
demonstrating that
an elastic domain, if it exists
at all, is smaller than the strain increment used in the simulations.
Irreversible strain occurred in directions tangent to the primary
yield surface, and the direction of the irreversible strain varied
with the direction of the stress increment.
For stress increments within
the deviatoric pi-plane,
the irreversible response had rounded-corners, evidence
of multiple plastic mechanisms.
The response at these
rounded corners varied in a continuous manner as a function
of stress direction.
The results are placed in the context of advanced
elasto-plasticity models:
multi-mechanism plasticity and tangential plasticity.
Although these models are an improvement
on conventional elasto-plasticity,
they do not fully fit the simulation results.
\end{abstract}
\begin{keyword}
  Granular material \sep
  plasticity \sep
  incremental response \sep
  stiffness \sep
  discrete element method
\end{keyword}
\end{frontmatter}
%
%
%
\section{\large Introduction}
The mechanical behavior of granular materials is exceedingly
complex, and engineers
have been challenged for decades
to measure, understand, and model this complexity.
The current work reveals further complexity, using
discrete element (DEM) simulations
to improve current understanding of the multi-directional
incremental
yield and flow characteristics of a material that is
initially loaded under monotonic triaxial conditions.
Although several frameworks have been developed for
structuring such experimental results, including
elasto-plasticity \cite{Yang:2005a},
hypoplasticity \cite{Wu:1996a,Lin:2015a},
generalized plasticity \cite{Hashiguchi:2005a},
damage plasticity \cite{Zhu:2010a},
micro-mechanics-based homogenization \cite{Nicot:2011c},
endochronic \cite{Yeh:2006a},
and shock and fracture-based \cite{Ganzenmuller:2011a} schema,
we will place our results in the
context of elasto-plasticity,
which is currently the prevailing framework for rate-independent
materials.
Conventional elasto-plasticity is founded on six principles:
\begin{enumerate}
\item
strain increments
can be separated into distinct
elastic and plastic parts as a sum or tensor
product 
with each part being
a homogeneous function of the stress increment,
and with the elastic increment being a reversible function
of the stress increment that is fully recovered after a closed
loading--unloading cycle
in stress-space;
\item
the elastic increment is a linear function~---
both homogeneous and additive~--- of the stress increment;
\item
the space of stress increments includes a finite elastic
region within which no plastic deformation occurs;
\item
the regions of elastic and plastic behavior are separated
by a hyperplane (incremental yield surface) in stress-space;
\item
plastic strain increments occur in a single direction,
which can depend upon the current stress and its history
but not upon the direction of the stress increment;
and
\item
the magnitude of the plastic strain is proportional
to the projection of the stress increment onto the
normal of the yield surface.
\end{enumerate}
These principles have been tested with both laboratory experiments
and simulations, in which soils or virtual
assemblies of particles are loaded in small probes
of stress or strain.
Four laboratory programs
and eleven simulation studies
are summarized in Table~\ref{table:studies}.
\begin{table}
  \centering
  \small
  \caption{Summary of multi-directional probe studies
           with laboratory tests and numerical simulations.
           \label{table:studies}}
  \begin{tabular}{lccccc}
  \toprule
    &&&& Probe & \\
    Study & Shape & Contacts & Method\textsuperscript{1} &
    type\textsuperscript{2} & $|d\boldsymbol{\varepsilon}|$ or $|d\boldsymbol{\sigma}|$\\
  \midrule
    Lewin \& Burland \cite{Lewin:1970a} &
    shale powder & triaxial & --- & --- & 10~kPa\\
    Tatsuoka \& Ishihara \cite{Tatsuoka:1974b}&
    sand & triaxial & --- & E--P & 0.001--0.006 \\
    Bardet
    \cite{Bardet:1994a} &    disks &   linear & DEM & E--P & $\approx 1$$\times$$10^{-3}$\\
    Anandarajah et al. \cite{Anandarajah:1995a} &
      sand & --- & triaxial & E--P & $\approx 1$$\times$$10^{-4}$ \\
    Royis \& Doanh \cite{Royis:1998a} &
      sand & --- & triaxial & E--P & $\approx 1$$\times$$10^{-4}$\\
    Kishino
    \cite{Kishino:2003a} &   spheres & linear & GEM & E--P & 1~kPa\\ 
    Alonso-Marroqu\'{i}n
    \cite{AlonsoMarroquin:2005a,AlonsoMarroquin:2005b} &
                             polygons & linear & DEM & E--P & 10~kPa \\
                             &&&&&$\approx 1$$\times$$10^{-5}$\\
    Calvetti et al.
    \cite{Calvetti:2003a,Tamagnini:2005a} &
                             spheres & linear$^{3}$ & DEM & R--I & $\approx 2$$\times$$10^{-4}$ \\
    Sibille et al.
    \cite{Sibille:2007a} &   spheres & linear &       DEM & R--I & --- \\
    Plassiard et al.
    \cite{Plassiard:2009a} & spheres & linear$^{4}$ & DEM & R--I & $\approx 2$$\times$$10^{-5}$\\
    Froiio and Roux
    \cite{Froiio:2010a} &    disks &   linear &  DEM\textsuperscript{5} & --- & $\approx 4$$\times$$10^{-6}$\\
    Harthong \& Wan
    \cite{Harthong:2013a} &  spheres & linear &       DEM & E--P & 0.1kPa \\
    Wan \& Pinheiro
    \cite{Wan:2014a} &       spheres & --- &          DEM & R--I \& E--P & 0.1~kPa \\
    Kuhn \& Daouadji \cite{Kuhn:2018a}&
    sphere-clusters & linear \& Hertz & DEM & R--I \& E--P & $2$$\times$$10^{-6}$\\
     Current work & sphere-clusters &Hertz &DEM &R--I & $2$$\times$$10^{-6}$\\
  \midrule
    \multicolumn{5}{l}{$^{1}\,$ GEM = Granular Element Method;
                                triaxial = laboratory experiments}\\
    \multicolumn{6}{l}{{}$^{2}\,$
                       R--I = reversible-irreversible;
                       E--P = elastic-plastic (see Section~\ref{sec:probes})}\\
    \multicolumn{5}{l}{$^{3}\,$ with rotational restraint of particles}\\
    \multicolumn{5}{l}{$^{4}\,$ with rolling friction of contacts}\\
    \multicolumn{5}{l}{$^{5}\,$ hybrid method: assembly loaded with DEM;
                                probes applied with method similar
                                to GEM \cite{Kishino:2003a,Agnolin:2007c}}\\
  \bottomrule
  \end{tabular}
\end{table}
The simulation studies include the early
two-dimensional (2D) studies of
Bardet \cite{Bardet:1994a} and
Alonso-Marroqu\'{i}n \cite{AlonsoMarroquin:2005b}
and more recent three-dimensional (3D)
simulations using sphere assemblies.
Although laboratory tests and simulations have exposed important
aspects of granular behavior,
disagreement or ambiguity still remains some
the principles enumerated above.
Tables~\ref{table:2Dprinciples} and~\ref{table:3Dprinciples}
summarize simulation results for 2D and
3D probe studies.
\begin{table}
  \centering
  \small
  \caption{Results of previous 2D simulations and their conformance
           with the six principles of conventional elasto-plasticity:
           Y = conforms with the principle, N = contradicts the
           principle.
           \label{table:2Dprinciples}}
  \begin{tabular}{lccccc}
    \toprule
    & & Alonso-Marroqu\'{i}n & Froiio \& & Sibille & Plassiard \\
    Elasto-plasticity principle & Bardet \cite{Bardet:1994a} &
    et al. \cite{AlonsoMarroquin:2005a,AlonsoMarroquin:2005b}
    & Roux \cite{Froiio:2010a} & et al. \cite{Sibille:2007a}$^{\ast}$
    & et al. \cite{Plassiard:2009a}$^{\ast}$\\
    \midrule
    (1)~$d\boldsymbol{\varepsilon}=
    d\boldsymbol{\varepsilon}^{\text{(e)}}+d\boldsymbol{\varepsilon}^{\text{(p)}}$,
    $d\boldsymbol{\varepsilon}^{\text{(e)}}$ is reversible &&Y& & & \\
    (2)~$d\boldsymbol{\varepsilon}^{\text{(e)}}$ linear: $d\varepsilon_{ij}^{\text{(e)}}=C_{ijkl}d\sigma_{kl}$ &Y&Y&&&Y\\
    (3)~Finite elastic domain &&Y&&Y&\\
    (4)~$d\boldsymbol{\varepsilon}^{\text{(e)}}$ \&
        $d\boldsymbol{\varepsilon}^{\text{(p)}}$
        domains are semi-spaces, normal $\mathbf{f}$ &Y&Y&Y&Y&Y\\
    (5)~Plastic increments $d\boldsymbol{\varepsilon}^{\text{(p)}}$
        in single flow direction $\mathbf{g}$ &Y&Y&Y&Y&Y\\
    (6)~ $|d\boldsymbol{\varepsilon}^{\text{(p)}}|= \mathbf{f}\cdot d\boldsymbol{\sigma}$ &Y&Y&Y&&Y\\
    \midrule
    \multicolumn{6}{l}{$^{\ast}$ Three-dimensional assemblies, but with
                       axisymmetric triaxial probes.}\\
    \bottomrule
  \end{tabular}
\end{table}
\begin{table}
  \centering
  \small
  \caption{Results of previous 3D simulations and their conformance
           with the six principles of conventional elasto-plasticity:
           Y = conforms with the principle, N = contradicts the
           principle.
           \label{table:3Dprinciples}}
  \begin{tabular}{lccccc}
    \toprule
    & Kishino & Calvetti & Tamagnini & Harthong & Wan \&\\
    Elasto-plasticity principle & \cite{Kishino:2003a} &
    et al. \cite{Calvetti:2003a}
    & et al. \cite{Tamagnini:2005a}
    & \& Wan \cite{Harthong:2013a}& Pinheiro \cite{Wan:2014a}\\
    \midrule
    (1)~$d\boldsymbol{\varepsilon}=
    d\boldsymbol{\varepsilon}^{\text{(e)}}+d\boldsymbol{\varepsilon}^{\text{(p)}}$,
    $d\boldsymbol{\varepsilon}^{\text{(e)}}$ is reversible
        & & & & & Y \\
    (2)~$d\boldsymbol{\varepsilon}^{\text{(e)}}$ linear: $d\varepsilon_{ij}^{\text{(e)}}=C_{ijkl}d\sigma_{kl}$
        & Y & Y & & &\\
    (3)~Finite elastic domain
        & Y & Y & N & & \\
    (4)~$d\boldsymbol{\varepsilon}^{\text{(e)}}$ \&
        $d\boldsymbol{\varepsilon}^{\text{(p)}}$
        domains are semi-spaces, normal $\mathbf{f}$
        & N & Y$^{\ast}$ &  & N &\\
    (5)~Plastic increments $d\boldsymbol{\varepsilon}^{\text{(p)}}$
        in single flow direction $\mathbf{g}$
        & N & & N & N & N \\
    (6)~$|d\boldsymbol{\varepsilon}^{\text{(p)}}|= \mathbf{f}\cdot d\boldsymbol{\sigma}$
        & N & Y & & & \\
    \midrule
    \multicolumn{6}{l}{$^{\ast}$ ``Y'' applies to virgin loading conditions.
    A finite elastic domain was not found with pre-loaded conditions.}\\
    \bottomrule
  \end{tabular}
\end{table}
The 2D simulations in Table~\ref{table:2Dprinciples}
are limited in the range of stress-space
that can be accessed, and in this sense, they are similar
to 3D axisymmetric triaxial conditions, which limit the accessible
space to a two-dimensional hyperplane of the principal stress components,
and the 3D studies also shown in the same table.
This table shows that, when tested,
each of the six elasto-plasticity principles is affirmed with
2D simulations, as indicated by the ``Y'' cells.
Some behaviors were not tested in certain studies,
and these cells are left blank.
Conspicuous ambiguity arises, however, in
the results of 3D simulations conducted with true-triaxial
conditions, in which the increments of
all three principal stresses were independently varied
(Table~\ref{table:3Dprinciples}).
For example, Tamagnini et al. \cite{Tamagnini:2005a} found
that some plastic deformation, albeit small, occurred regardless of the
direction of small loading probes, a result that is
contrary to other 3D (and 2D) studies and to principle~3.
In particular, they found
that plastic strains occur for stress increments
in opposite directions
(i.e., for both loading and unloading),
thus violating the third principle.
Based upon their triaxial tests of sand,
Royis and Doanh \cite{Royis:1998a} also doubted the
existence of a finite elastic regime, as they detected
small plastic strain increments for all loading directions.
Yet principle~3 was affirmed in several other studies.
\par
In regard to principle~4,
some studies have found that the
elastic and plastic regions in incremental
stress-space 
are distinct and are separated by a flat hyperplane
(i.e., a yield surface without corners and
having unit normal direction $\mathbf{f}$)
\cite{Bardet:1994a,AlonsoMarroquin:2005a,Plassiard:2009a,%
Froiio:2010a,Calvetti:2003a,Sibille:2007a}.
Other programs, however, found evidence of a possibly cornered
yield surface \cite{Kishino:2003a,Tamagnini:2005a,Harthong:2013a}.
Similar ambiguity is found with the assumption of a single
direction of the plastic strain increment at a particular
stage of loading (principle~5).
In particular,
Kishino \cite{Kishino:2003a} and
Wan and Pinheiro \cite{Wan:2014a}
shed doubt on the fifth principle
with simulations that probed a material in multiple stress directions
and discovered small changes in the directions of the
resulting plastic strain increments.
The sixth principle, which disallows plastic strain
for stress increments that lie along the yield surface
(i.e., tangential increments),
was refuted by Kishino \cite{Kishino:2003a} and
Plassiard et al. \cite{Plassiard:2009a}, who detected small
plastic strains for these tangential increments.
The conclusions of these past studies are influenced,
in part, by the size of their stress or strain increments,
since larger increments can obscure the existence of a
cornered yield surface, the nature of the elastic domain,
the character of the plastic flow directions, and the effect of
a stress increment's direction on the plastic
strain increment.
We also note each past study
had its own particular focus, such that some of the aspects noted in
Tables~\ref{table:2Dprinciples} and~\ref{table:3Dprinciples}
are based upon limited data.
\par
The current work examines each of the six principles
with extensive and carefully controlled
simulations that employ extremely small
strain increments, a more realistic particle shape,
and a rigorous implementation of Hertzian contact between
particles, and
we present results that more fully characterize the incremental
response of a granular material~--- in our case, a virtual sand.
We show that only the second principle survives scrutiny,
yet even the linear form of the elastic response,
as implied with principle~2, develops
a rather extreme anisotropy that favors elastic dilation.
With each principle, we more fully ascertain
granular behavior and the nature of the nonconformity or agreement.
The first principle was the focus of a second article
by the authors \cite{Kuhn:2018a},
which demonstrated that elastic strain increments,
although recoverable, are not reversible,
a finding that is more fully described below
in the context of elastic-plastic coupling
\cite{Hueckel:1976a,Dafalias:1977b}.
As such, the elastic--plastic partition of strain
increments~--- the basis of
the first principle~--- is replaced with a reversible--irreversible
partition, and the terms ``reversible'' and ``irreversible'',
defined below, will
replace elastic and plastic in this work
(see also \cite{Hueckel:1977a,Collins:1997a}).
\par
Because the simulations demonstrate
a lack of conformity with five of the six principles,
we also consider more advanced elasto-plastic models,
which
allow a softer
response in certain directions and restrict the zone
of elastic behavior:
generalized plasticity
\cite{Pastor:1986a},
multi-mechanism plasticity \cite{Hill:1967a}, and
tangential plasticity \cite{Hashiguchi:2005a,Khojastehpour:2006a}.
These models were originally developed to explain
the susceptibility of materials to certain localized failure modes,
and their success in modeling our simulation
results is quite remarkable, considering that
they were originally based upon very sparse evidence of
the behavior of these materials.
\par
The plan of the paper is as follows.
In the following section, we review the application of discrete
element simulations for modeling incremental behavior,
present the particular features of the model
used in the current work, and discuss the alternative
separations of strain increments with
elastic--plastic and reversible--irreversible partitions.
In Section~\ref{sec:probes2}, we review our results of several
series of stress probes that test the six principles
of elasto-plastic materials set forth above.
The nature of the measured reversible and irreversible
strains are then critically examined
in Sections~\ref{sec:reversible}
and~\ref{sec:irreversible}.
Because of inconsistencies between the simulation results
and conventional elasto-plastic theory,
we assess two modest extensions of the theory~--- multi-mechanism and
tangential plasticity~--- and evaluate their fit
with the simulations.
%
%
\section{DEM Modeling}\label{sec:DEMmodeling}
DEM simulations have become an important means for characterizing
the directional yield and flow properties for small increments
of loading
and to decompose the resulting strain
increments into reversible and irreversible (or elastic and plastic) parts.
Simulations, both past and current,
are meant to augment understanding of
granular materials
gained from physical tests.
However, because micro-scale information and more
general boundary conditions are accessible with
DEM simulations, they can expose granular behavior in ways
that are not easily achieved with laboratory testing.
Specifically,
three advantages of DEM simulations were
exploited in our simulations.
First,
after creating a granular assembly and loading it
monotonically to
a given state, this state and all of its micro-data
are saved.
This saved state~--- which Alonso-Marroqu\'{i}n \cite{AlonsoMarroquin:2005b}
calls a numeric ``clone''~--- constitutes a consistent
reference state for subsequent loading probes. 
Second, in performing the loading probes,
the DEM code permitted control of arbitrary
combinations of six components of the
stress and strain tensors or any six linearly independent
combinations of stress
and strain~--- loading
conditions that would require multiple sample preparations
and even different testing apparatus in a physical laboratory
\cite{Kuhn:2002b}.
%
Finally, by using different variations of probes,
described in Section~\ref{sec:probes},
we measured the separate reversible and irreversible
constituents of the total strain
increments.
Because of these and other advantages, simulations have begun to supplant
physical experiments for investigating certain aspects of granular
behavior. 
\subsection{DEM model and monotonic triaxial compression}\label{sec:monotonic}
We used a series of slow, quasi-static
DEM simulations to study the multi-directional incremental response of
a granular assembly
(model details are described in \ref{sec:DEM}).
Unlike the dynamic regime of behavior,
in which shock waves can move through the material
\cite{Steinhauser:2009a,Chakraborty:2013a},
our slow strain rate assures that particles
remain in near-equilibrium and that results are independent of
the loading rate.
The assembly was a cubical box filled with 10,648
smooth non-convex sphere-cluster particles contained within
periodic boundaries.
The particles' non-convex shapes were intended to capture the irregular
shapes of sand particles, which permit the nestling of particle
pairs
that share multiple contacts.
The shape, sizes, and arrangement of the particles were
calibrated to closely simulate the behavior of the
fine-grain poorly-graded medium-dense Nevada Sand
(see the Appendix and \cite{Kuhn:2014c}). 
The assembly was large enough to
capture the average material behavior but sufficiently small to
prevent meso-scale localization, such as shear bands
\cite{Kuhn:2009b,Wan:2014a}.
A distinctive feature of sands is the dependence
of the elastic moduli upon the
confining (mean) stress.
To capture this behavior, the
contacts between particles were modeled with a Hertz normal
response and a full implementation of the Cattaneo--Mindlin
tangential response,
which permits slip, micro-slip, and elastic behaviors
at the contacts \cite{Mindlin:1953a,Kuhn:2011a}.
To model such non-linear contact behavior,
most DEM studies use
an approximation of the Cattaneo--Mindlin contact,
with the tangential stiffness taken
as a multiple of the normal stiffness, which changes with
the normal force.
This common approximation can produce an unintended
infusion of energy during a closed cycle of contact loading
\cite{Elata:1996a}.
Because we relied upon loading-unloading probes,
we deemed an energetically-consistent
full implementation of the Cattaneo--Mindlin contact
as essential for determining
the incremental stiffness characteristics of granular materials
(\ref{sec:DEM}).
The dense initial particle arrangement was isotropic with a confining
(negative mean)
stress of 100~kPa and porosity 0.363 (void ratio of 0.570).
During subsequent loading, the inter-particle friction coefficient
was 0.55.
\par
Although it is inconsistent with the quasi-static hypothesis,
DEM results can depend upon
the loading rate~--- an unfortunate dependence which is infrequently
discussed in a candid manner in the DEM literature.
(Such rate effects have, for decades,
been recognized in laboratory soil testing, 
and measures are usually taken to reduce these effects,
e.g.,
\cite{Lewin:1970a}.)
To mitigate such rate dependence,
a slow loading rate and other measures
were taken, assuring that the simulations were nearly
quasi-static.
Because of their importance in the current study,
computational essentials are detailed in \ref{sec:DEM}, 
which also
describes quantitative indicators verifying the
rate-independent, quasi-static nature
of the simulations.
\par
\begin{figure}
  \centering
  \includegraphics{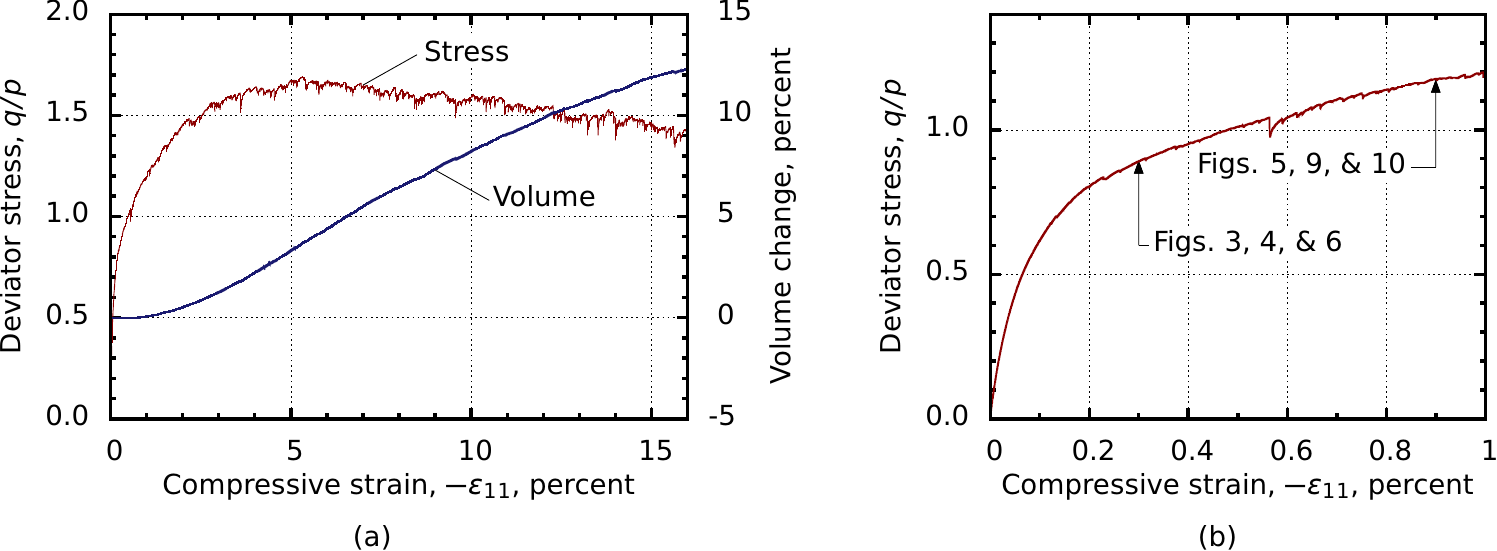}
  \caption{Stress, strain, and volume change during DEM simulation of
           triaxial compression
           at a constant mean stress of 100~kPa.
           \label{fig:StressStrain}}
\end{figure}

The loading conditions in this study were orthotropic
with no transverse shearing, so that
the principal stress and strain rates did not rotate during
deformation, obviating the need to resolve non-coaxial
behavior
or distinguish between a co-rotated
stress rate and the simpler Cauchy rate
$\dot{\boldsymbol{\sigma}}$.
Figure~\ref{fig:StressStrain} shows the results of
drained isobaric (constant-$p$) axisymmetric
triaxial compression,
in which the $x_{1}$ width
of the assembly was reduced at constant rate $\dot{\varepsilon}_{11}$,
while adjusting the lateral widths to maintain a constant mean stress
of 100~kPa.
The lateral stresses, $\sigma_{22}$ and $\sigma_{33}$, were maintained
nearly equal during this monotonic constant-$p$ compression,
with
($\sigma_{22}\approx\sigma_{33}$, see \ref{sec:DEM} for verification
of the careful control of stresses during monotonic loading and
stress probes).
In the figure, the deviator stress $q$ is the negative of
$\sigma_{11}-\tfrac{1}{2}(\sigma_{22}+\sigma_{33})$,
and pressure (negative mean stress)
$p=-\tfrac{1}{3}(\sigma_{11}+\sigma_{22}+\sigma_{33})$.
All of the subsequent stress probes began from initial states
taken along
this path of monotonic axisymmetric loading.
\subsection{Stress probe methods}\label{sec:probes}
The separation of strain increments into elastic and plastic parts is
an established concept in constitutive modeling,
and the part of a strain increment
that is recovered after completing a
closed loading--unloading cycle in stress-space is usually assumed to be
free of contact slip;
whereas, the remaining, permanent part of a strain increment
is usually associated
with frictional dissipation during such closed cycles.
Collins and Houlsby noted, however,
that recoverable (elastic) strain increments
might not be entirely dissipation-free (reversible),
and permanent (plastic) strain increments
might not be purely dissipative (irreversible, see also
\cite{Nicot:2006a}).
Hueckel \cite{Hueckel:1976a,Hueckel:1977a,Hueckel:1979a}
noted that the micro-structure
of a material can be altered by internal changes that accompany
plastic deformation, and that these internal changes
produce modifications of the elastic moduli concurrently with the loading
(termed \emph{elastic-plastic coupling}).
These internal changes,
tracked perhaps with internal variables \cite{Dafalias:1977b},
accompany dissipation but can be caused by processes that are,
themselves, dissipation-free.
Pouragha and Wan \cite{Pouragha:2017a,Pouragha:2018a} have noted that the creation
and disintegration of contacts and geometrical alterations
of the contact arrangement, which occur during deformation, are such
non-dissipative mechanisms that can alter the elastic moduli with
loading.
The distinction between elastic and reversible
strain increments must, therefore,
be considered when planning incremental probes,
as different DEM techniques are available for
determining the elastic--plastic and reversible--irreversible partitions
of a strain increment.
The authors have used both techniques and found small differences in
the resulting partitions,
demonstrating the existence of elastic--plastic coupling
and contradicting the principle~1 of conventional elasto-plasticity
\cite{Kuhn:2018a}.
These DEM results differ from those of Wan and Pinheiro
\cite{Wan:2014a}, who found an equivalence of the
reversible and elastic strains.
Although the reversible--irreversible partition can not be
ascertained in a laboratory setting,
this partition is considered more
fundamental, and it was shown to produce a more regular separation of
the incremental response
(for example, a linear stiffness relationship between
increments of stress and reversible strain).
\par
\begin{figure}
  \centering
  \includegraphics{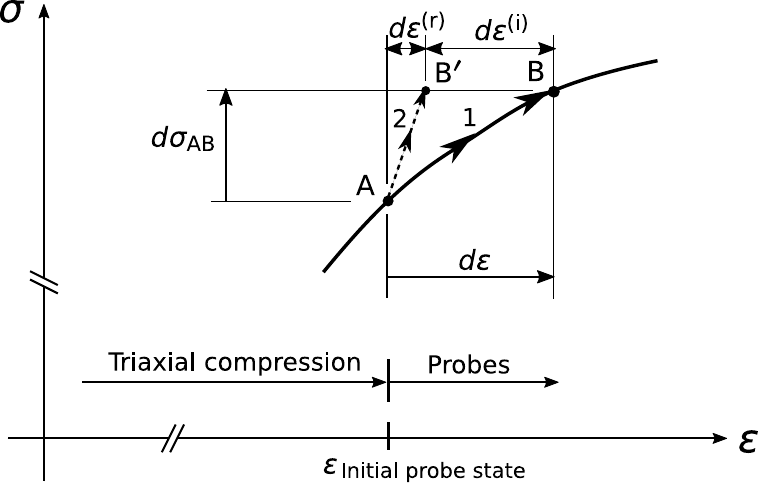}
  \caption{Probes AB and AB$^{\prime}$ to determine the
           total and reversible strain increments.%
           \label{fig:probes}}
\end{figure}

For these reasons, we have measured
the reversible
and irreversible strains~--- in both
magnitude and direction~--- by
using a DEM technique pioneered by Calvetti and his co-workers
\cite{Tamagnini:2005a}.
The initial, reference states for our series of probes were
first established
during the course of
drained monotonic axisymmetric isobaric
(constant-$p$) triaxial compression
(Fig.~\ref{fig:StressStrain}).
At several reference strains,
we fully saved the status of all particles and contacts,
so that different deformation probes could be started from
these ``initial probe states''
(point~A in Fig.~\ref{fig:probes}).
Starting at an initial probe state,
each probe was begun by
advancing strain in a particular direction
to produce a strain increment of magnitude
$|d\boldsymbol{\varepsilon}|=2\times 10^{-6}$
(the increment~AB in Fig.~\ref{fig:probes}),
where the norm
$|d\boldsymbol{\varepsilon}|
=\sqrt{d\varepsilon_{11}^2+d\varepsilon_{22}^2+d\varepsilon_{33}^2}$.
At the end of this probe
(point B), we noted the resulting increments
of all strain and stress components during this initial probe.
To determine the reversible part of this deformation,
we conducted a dissipation-free ``reversible probe'' (AB$^\prime$
in Fig.~\ref{fig:probes})
by replicating the same \emph{stress increment} as during
loading (the increment
$d\boldsymbol{\sigma}_{\text{AB}}$)
but computationally preventing contact slip
(and micro-slip)
by assigning a large friction coefficient (i.e., $\mu=50$).
That is,
we first determined the stress increment
$d\boldsymbol{\sigma}_{\text{AB}}$ that resulted
from unadulterated loading with a contact friction
coefficient $\mu=0.55$,
and then we produced this same stress increment
with $\mu=50$.
The latter strain increments
were free  of contact slip and were entirely the result of elastic contact
movements: these strain increments are the reversible
strains $d\boldsymbol{\varepsilon}^{\text{(r)}}$,
and the irreversible strains $d\varepsilon_{ij}^{\text{(i)}}$
are the difference
$d\varepsilon_{ij}-d\varepsilon_{ij}^{\text{(r)}}$.
We note, again, that the reversible strain is not the
``recoverable strain,'' as irreversible processes
in the form of contact slip, contact micro-slip,
and particle rearrangement
occur at all stages and in all directions of a loading--unloading
cycle in stress-space.
The technique for isolating the reversible part
of a strain increment was used in \cite{Kuhn:2018a}
to demonstrate that granular materials exhibit
elastic--plastic coupling and
that during strain hardening,
contact slip~--- a distinctly irreversible mechanism~---
accompanies loading in any direction.
Another irreversible mechanism, the non-dissipative gain
and loss of contacts, can also occur during deformation \cite{Pouragha:2017a},
and this mechanism was not precluded in the AB$^{\prime}$ probes.
As such, our technique of measuring the reversible--irreversible
partition involves an unavoidable approximation.
We note, however, that contact loss and gain was rarely
observed during the strain increments of $2\times 10^{-6}$.
\section{\large Multi-direction stress probes}\label{sec:probes2}
Stress and strain probes were used to characterize
yield and flow
for loading in multiple directions,
a purpose for which
the advantages of DEM simulations
are most persuasive. 
The current work extends the studies in Table~\ref{table:studies} by
conducting a comprehensive series of probes that
cover the three-dimensional space of incremental
principal stress.
We also submit the results to a close scrutiny that
reveals new phenomena and insights of the elasto-plastic behavior
of granular materials.
The probe strains are of size $2\times 10^{-6}$,
much smaller than previous studies, and
the current simulations differ in several other
respects from past studies:
the use of non-convex sphere-clusters,
an exact Hertz-like Cattaneo--Mindlin contact model,
and the slow loading necessary to produce quasi-static conditions.
\subsection{Framework of generalized stress and strain}\label{sec:generalized}
We conducted several series of
incremental probes in the three-dimensional
space of normal strains,
$[\boldsymbol{\varepsilon}]=[\varepsilon_{11},\varepsilon_{22},\varepsilon_{33}]^{\mathsf{T}}$,
and in the complementary space
of normal stresses,
$[\boldsymbol{\sigma}]=[\sigma_{11},\sigma_{22},\sigma_{33}]^{\mathsf{T}}$,
where extensional strains and tensile stresses are positive.
Because the 3D rectangular assembly was not sheared
and only rectilinear deformations were applied,
these normal components are the principal strains and stresses that
fully express the deformation and loading conditions.
An assembly was first loaded in constant-$p$ triaxial compression
to an initial probe state (Fig.~\ref{fig:probes}),
and the subsequent probes were true-triaxial increments of
the strain and stress increments,
$[d\boldsymbol{\varepsilon}]$ and $[d\boldsymbol{\sigma}]$.
Rather than express the results in Lode coordinates or principal
strain coordinates,
we normalize and plot the deviatoric and
volumetric components of these increments in a systematic
manner, by transforming the Cartesian components
of strain into an alternative
set of generalized components, for which
the unit vectors $\vec{\mathbf{e}}_{1}$, $\vec{\mathbf{e}}_{2}$, and
$\vec{\mathbf{e}}_{3}$ serve as an orthonormal basis,
organized as the columns of a matrix $[\mathbf{E}]$,
\begin{equation}\label{eq:Esystem}
  [\mathbf{E}]
  =
  \begin{bmatrix}
    \vec{\mathbf{e}}_{1},\; \vec{\mathbf{e}}_{2},\; \vec{\mathbf{e}}_{3}
  \end{bmatrix}
  =
  \begin{bmatrix}
  \displaystyle\frac{-1}{\sqrt{3}}
    \begin{bmatrix}
      1\\1\\1
    \end{bmatrix}, &
  \displaystyle\frac{-1}{\sqrt{2}}
    \begin{bmatrix}
      0\\1\\-1
    \end{bmatrix}, &
  \displaystyle\frac{-1}{\sqrt{6}}
    \begin{bmatrix}
      2\\-1\\-1
    \end{bmatrix}
  \end{bmatrix}
\end{equation}
%
When represented in the three-dimensional space
of Cartesian (principal)
strains, $\varepsilon_{11}$, $\varepsilon_{22}$, and
$\varepsilon_{33}$, these
basis vectors have the following meanings:
$\vec{\mathbf{e}}_{1}$ is the
compressive volumetric direction; 
$\vec{\mathbf{e}}_{2}$ is a transverse-deviatoric direction
perpendicular to the
Rendulic plane (i.e., perpendicular to the plane of volume strain $v$
and strain $\varepsilon_{11}$);
and $\vec{\mathbf{e}}_{3}$ is the deviatoric direction 
that would be accessible by axisymmetric triaxial conditions,
within the Rendulic plane.
Because matrix $[\mathbf{E}]$ is orthonormal, the conventional
strain $[\boldsymbol{\varepsilon}]$ and stress
$[\boldsymbol{\sigma}]$ can be scaled
and expressed in the new basis vectors
with the generalized scalar components $[e_{1}, e_{2}, e_{3}]^{\mathsf{T}}$
and $[s_{1}, s_{2}, s_{3}]^{\mathsf{T}}$:
\begin{align}\label{eq:transform}
  [\boldsymbol{\varepsilon}] &=
  e_{1}\vec{\mathbf{e}}_{1} + e_{2}\vec{\mathbf{e}}_{2} + e_{3}\vec{\mathbf{e}}_{3}
  ,\quad
  [\boldsymbol{\varepsilon}] =
  [\mathbf{E}][\mathbf{e}]
  ,\quad\text{and} \quad
  [\mathbf{e}]=[\mathbf{E}]^{\mathsf{T}}[\boldsymbol{\varepsilon}]\\
  \label{eq:Ssystem}
  [\boldsymbol{\sigma}] &=
  s_{1}\vec{\mathbf{s}}_{1} + s_{2}\vec{\mathbf{s}}_{2} + s_{3}\vec{\mathbf{s}}_{3}
  ,\quad
  [\boldsymbol{\sigma}] =
  [\mathbf{E}][\mathbf{s}]
  ,\quad\text{and} \quad
  [\mathbf{s}]=[\mathbf{E}]^{\mathsf{T}}[\boldsymbol{\sigma}]
\end{align}
%
(note the lack of an over-arrow with the scalar lists
$[\mathbf{e}]$ and $[\mathbf{s}]$).
Henceforth,
we refer to the 1--2--3 components as volumetric,
transverse-deviatoric, and deviatoric, respectively,
and we refer to the Rendulic plane,
the pi-plane, and the transverse-orthogonal plane as
planes with $\mathbf{e}_{2}=0$, $\mathbf{e}_{1}=0$,
and $\mathbf{e}_{3}=0$, respectively
\cite{Rendulic:1936a}.
Note that volume strain $e_{1}$ and isotropic stress
$s_{1}$ are compressive.
\par
The term ``tangential'' has been applied to stress increments
that are orthogonal to the \emph{normal of the yield surface}
(direction $\mathbf{f}$ in Section~\ref{sec:irreversible})
\cite{Hashiguchi:2005a}.
For the axisymmetric condition of our initial triaxial compression,
the yield surface normal should lie within the
$\vec{\mathbf{s}}_{1}$--$\vec{\mathbf{s}}_{3}$ plane
(the Rendulic plane),
and the $\vec{\mathbf{s}}_{2}$ direction
should be orthogonal to (tangential to) the yield normal.
The full tangent plane will also include components
in the $\vec{\mathbf{s}}_{1}$ and $\vec{\mathbf{s}}_{3}$
directions.
%
%
The generalized stresses and strains
of Eqs.~(\ref{eq:transform})--(\ref{eq:Ssystem})
are work-conjugate and
are particularly convenient
for planning, plotting, and analyzing the results of
simulation probes.
%
%
%
\subsection{Results of strain and stress probes}\label{sec:ProbeResults}
\begin{figure}
  \centering
  \includegraphics{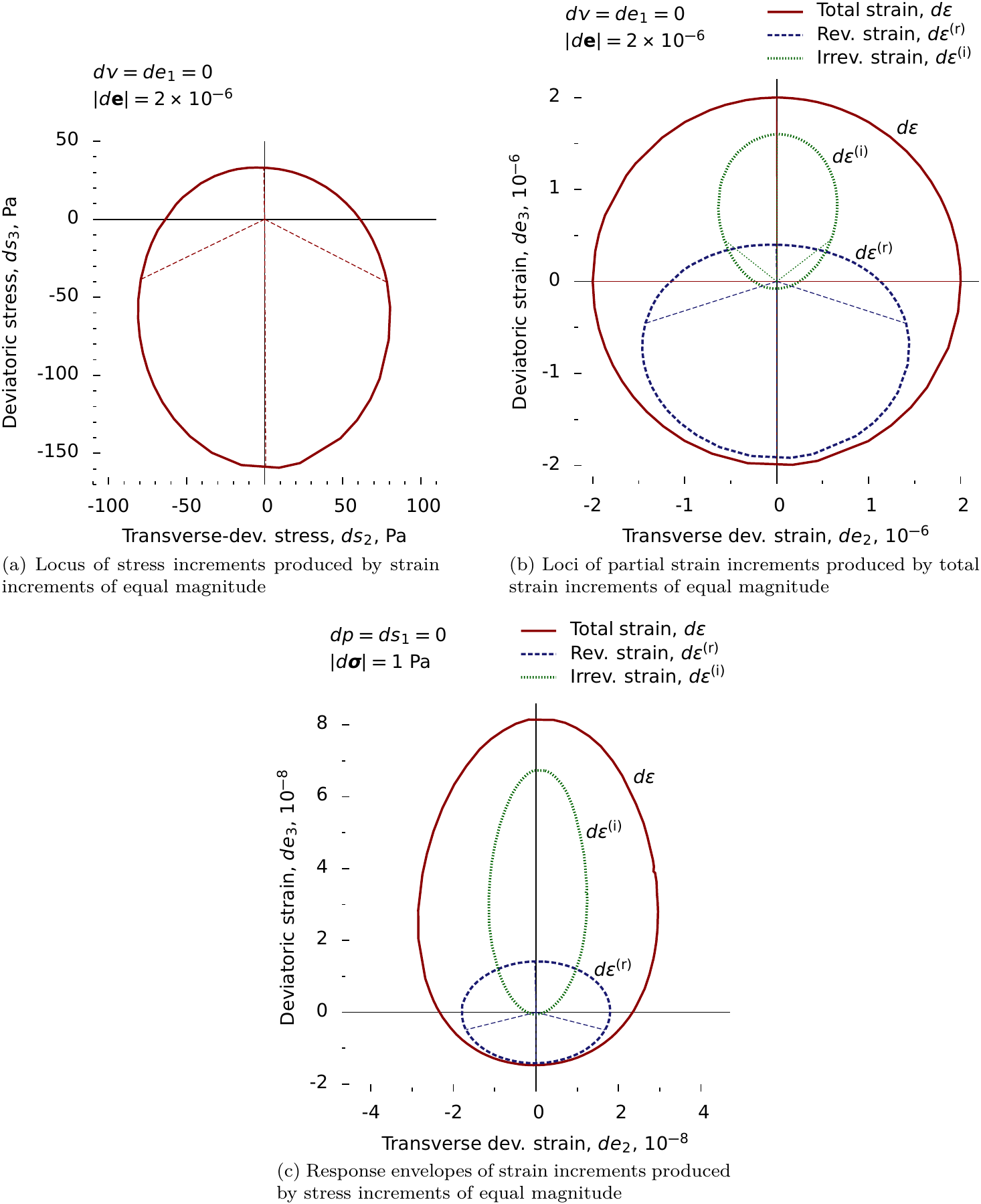}
%
  \caption{Pi-planes of deviatoric strain-probes,
           conducted at the reference strain
           $\varepsilon=-0.3\%$:
           (a) locus of stress increments produced with
           deviatoric strain increments
           of equal magnitude 
           under undrained, isochoric conditions;
           (b) loci of reversible and irreversible strain parts,
           for total strain
           increments of equal magnitude 
           under undrained, isochoric conditions; and
           (c) response envelopes of 
           strain increments that resulted from deviatoric
           stress increments of equal magnitude
           under isobaric (drained, constant-$p$) conditions.
           \label{fig:piplane}}
\end{figure}

In this section,
we describe the primary trends of the reversible and irreversible
strain increments that were measured with probes that started from
an initial stage
of constant-$p$ triaxial compression (Fig.~\ref{fig:StressStrain}).
Three series of probes were conducted:
probes within each of
the three planes $\vec{\mathbf{e}}_{1}$--$\vec{\mathbf{e}}_{2}$,
$\vec{\mathbf{e}}_{1}$--$\vec{\mathbf{e}}_{3}$, and
$\vec{\mathbf{e}}_{2}$--$\vec{\mathbf{e}}_{3}$ of generalized
strain or stress.
Figure~\ref{fig:piplane} shows the results of deviatoric probes
in the octahedral pi-plane,
$\vec{\mathbf{e}}_{2}$--$\vec{\mathbf{e}}_{3}$,
for probes that were either isochoric
or isobaric,
and each sub-figure is a plane of incremental
strain, $de_{2}$ and $de_{3}$, or of incremental
stress, $ds_{2}$ and $ds_{3}$.
The conditions of this figure could be accessed,
for example, with
true-triaxial equipment in a laboratory setting.
Continued axisymmetric
triaxial compression is in the upward direction of these
figures,
and a reversal of triaxial compression
(unloading or triaxial extension)
is downward.
The probes in Fig.~\ref{fig:piplane} were conducted
during the early stage of strain hardening, at the reference strain
$\varepsilon_{11}=-0.3\%$, for which the deviator stress was about
55\% of the peak strength (see Fig.~\ref{fig:StressStrain}b).
Each sub-figure is the result of at least eighty probes.
\par
Figure~\ref{fig:piplane}a shows the magnitudes of the changes in stress
$d\boldsymbol{\sigma}$ that resulted from isochoric (undrained)
strain probes that shared a deviatoric strain magnitude,
$|d\boldsymbol{\varepsilon}|
=\sqrt{d\varepsilon_{11}^2+d\varepsilon_{22}^2+d\varepsilon_{33}^2}=2\times 10^{-6}$.
That is, Fig.~\ref{fig:piplane}a shows the locus of
stress increments
$[ds_{2},ds_{3}]^{\mathsf{T}}$
produced by strain probes with an equal Euclidean magnitude
of the deviatoric strain increment
$[de_{2},de_{3}]^{\mathsf{T}}$.
As expected, the behavior is softer (smaller radial distances)
in the direction
of continued triaxial compression (upward) than for
triaxial unloading (downward).
%
\par
Figure~\ref{fig:piplane}b gives additional results for
deviatoric
isochoric strain probes with an equal magnitude of strain
(radius 2$\times$10\textsuperscript{-6} of the total strain
circle).
The reversible and irreversible strains were determined 
with the method described in the previous section and illustrated
in Fig.~\ref{fig:probes},
and Fig.~\ref{fig:piplane}b gives the loci of the
irreversible and reversible strain increments that were
produced by probes of equal strain magnitude
$|d\boldsymbol{\varepsilon}|$.
The irreversible strain increment
$d\boldsymbol{\varepsilon}^{\text{(i)}}_{ij}$
was the largest part the total strain increment
when the probe was in the direction of continued
triaxial compression (upward in the figure); whereas,
reversible strain increment
$d\boldsymbol{\varepsilon}^{\text{(r)}}_{ij}$
was the largest part for triaxial
unloading (downward) and for neutral (horizontal) probes that
produced transverse-deviatoric strains in direction
$d\vec{\mathbf{e}}_{2}$.
A careful inspection of Fig.~\ref{fig:piplane}b
reveals that a small irreversible strain is also produced in the
downward ``unloading'' direction of triaxial extension,
a result that was also mentioned by Tamagnini et al.
\cite{Tamagnini:2005a} and is explored more fully below.
%
\par
Figure~\ref{fig:piplane}c shows
the reversible and irreversible
parts of
strain increments that were produced by deviatoric \emph{isobaric}
stress probes having
an equal incremental stress magnitude of 1~Pa
(i.e., a magnitude $|d\boldsymbol{\sigma}| = |d\mathbf{s}|=\sqrt{ds_{1}^2+ds_{2}^2+ds_{3}^2}=1$~Pa).
Each probe had a strain magnitude of $2\times 10^{-6}$,
and after noting the resulting stress increment,
the strain was scaled to the stress magnitude of 1~Pa.
Like the response envelopes of Gudehus \cite{Gudehus:1979a},
this figure displays the strain increments produced by
stress increments that share an equal magnitude
but are applied in different directions.
The irreversible response in Fig.~\ref{fig:piplane}c is not a straight line
but has a narrow elliptic shape, an indication that,
contrary to conventional elasto-plasticity (principle~5),
a single flow direction does not apply to granular materials.
Our results are similar to those of
Kishino \cite{Kishino:2003a},
Tamagnini et al. \cite{Tamagnini:2005a},
and Wan and Pinheiro \cite{Wan:2014a},
although the irreversible
response envelopes in these earlier studies
had a tear-drop shape
instead of the elliptic shape of our simulations
(see also \cite{Darve:2005b} for a discussion).
The dependence of flow direction on loading direction is more
fully described in Section~\ref{sec:irreversible}.
%
\par
\begin{figure}
  \centering
  \includegraphics{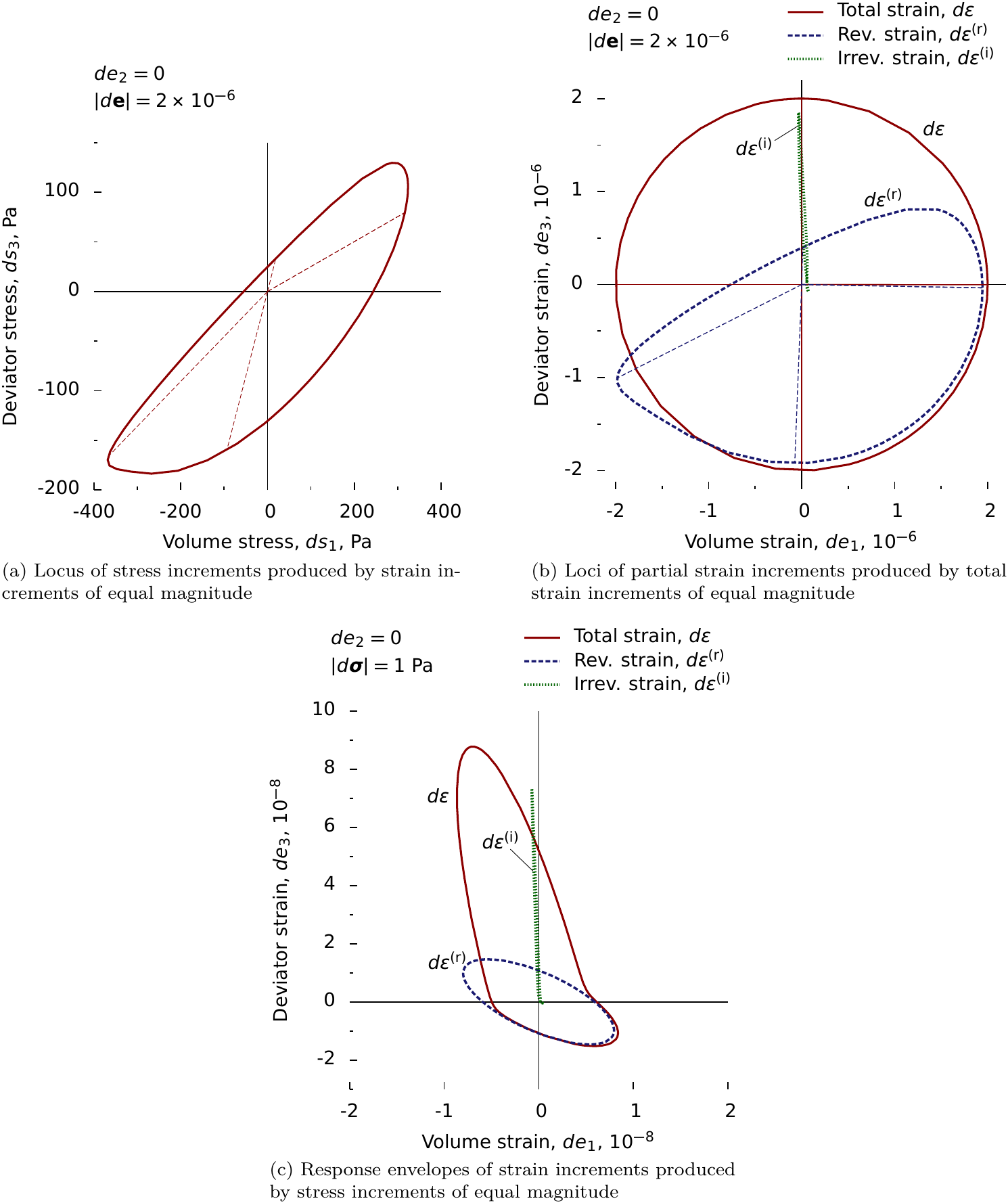}
%
  \caption{Results of strain probes in the
           Rendulic volumetric--deviatoric plane,
           conducted at the strain
           $\varepsilon_{11}=-0.3\%$:
           (a) locus of stress increments produced by strain increments
           of equal magnitude; 
           (b) loci of reversible and irreversible strain parts,
           for total strain
           increments of equal magnitude; and 
           (c) response envelopes of
           strain increments that resulted from
           volumetric--deviatoric
           stress increments of equal magnitude, 1~Pa.
           \label{fig:rendulic}}
\end{figure}

The incremental response to deviatoric and \emph{volumetric}
loading is shown in
Fig.~\ref{fig:rendulic} with Rendulic diagrams
of the incremental strains, $de_{1}$ and $de_{3}$, and of the incremental
stresses, $ds_{1}$ and $ds_{3}$.
The conditions depicted in this figure could
be accessed with conventional
axisymmetric triaxial loading equipment
(e.g., the studies in Table~\ref{table:2Dprinciples}).
As with the previous figure,
the initial probe state of this figure is at
a reference strain $\varepsilon_{11}=-0.3\%$. 
At this strain, the deviator stress was about 55\% of the peak strength,
and
the volumetric behavior was slightly dilative, as the neutral
condition of zero volume strain
had occurred at an earlier strain of 0.2\%.
In the figure,
continued triaxial compression is upward;
a reversal of triaxial compression (triaxial extension) is downward;
isotropic compression is to the right; and expansion (dilation) is to
the left.
\par
Radial distances in Fig.~\ref{fig:rendulic}a are the magnitudes of
the stress increments that were produced by equal magnitudes of strain
$|d\boldsymbol{\varepsilon}|=|d\mathbf{e}|=2$$\times$$10^{-6}$.
As expected, the behavior is stiffer for volumetric increments than for
deviatoric increments and is softer for continued deviatoric loading
(upward) than
for deviatoric unloading (downward).
Figure~\ref{fig:rendulic}b shows magnitudes of the
reversible and irreversible parts of the
strain increments that were produced by equal magnitudes of total
strain: each curve is a locus of increments,
$de_{1}$ and $de_{3}$,
for an equal magnitude of the \emph{total} strain increment
(over eighty probes are represented).
At the early stage of strain hardening
represented in this figure,
both reversible and irreversible strains are of comparable magnitude.
Note that the locus of reversible strain magnitude crosses that of the
total strain,
an unusual result that is consistent with the
triangle inequality,
$|d\boldsymbol{\varepsilon}^{\text{(r)}}|+
|d\boldsymbol{\varepsilon}^{\text{(i)}}|\ge
|d\boldsymbol{\varepsilon}|$.
\par
In Fig.~\ref{fig:rendulic}c, we show the deviatoric--volumetric
response envelopes for equal-magnitude \emph{stress increments},
$|d\boldsymbol{\sigma}| = |d\mathbf{s}|=1$~Pa.
Each probe had a strain magnitude of $2\times 10^{-6}$,
and after noting the resulting stress increment,
the strain was scaled to the stress magnitude of 1~Pa.
The elliptical shape of the response
envelope of reversible strain indicates a greater
reversible stiffness for volumetric than for deviatoric loading,
and the ellipse's tilt is evidence of a coupling of volumetric
and deviatoric strains through the Poisson effect.
The irreversible strain appears as a nearly straight line
that radiates from the origin, indicating a single direction of
the irreversible strain when viewed in the Rendulic plane,
a result
similar to the simulations in Table~\ref{table:2Dprinciples}
and the soil experiments of
Anandarajah et al. \cite{Anandarajah:1995a},
Royis and Doanh \cite{Royis:1998a}, and Darve and Nicot \cite{Darve:2005b}.
The irreversible strain
is predominantly deviatoric,
but with a slight tilt due to a small dilation tendency.
\par
\begin{figure}
  \centering
  \includegraphics{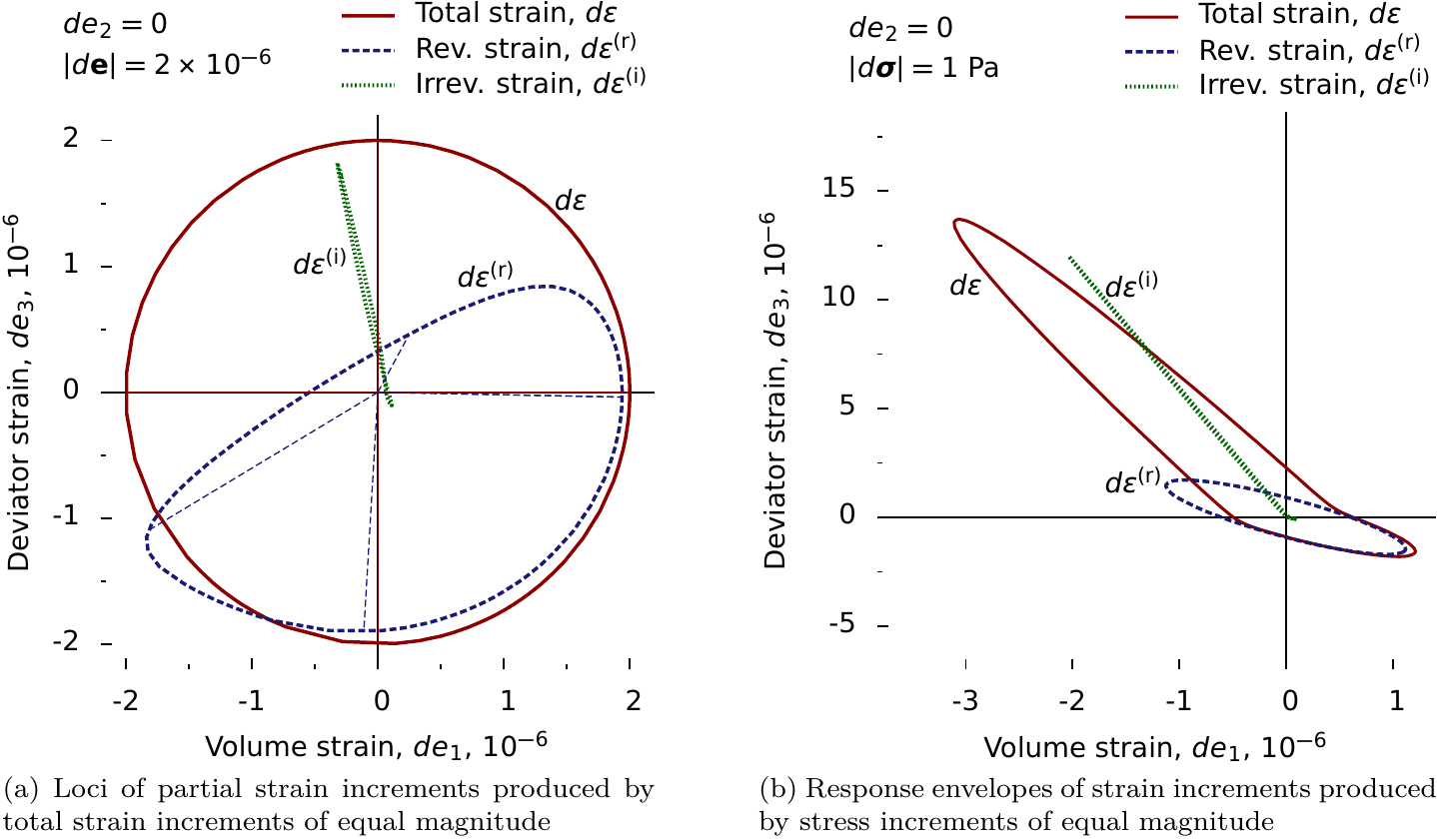}
%
  \caption{Results of strain probes in the
           Rendulic volumetric--deviatoric plane,
           conducted at the strain
           $\varepsilon_{11}=-0.9\%$:
           (a) loci of volumetric and deviatoric parts for total
           strain increments 
           of equal magnitude; and
           (b) response envelopes of
           strain increments that resulted from
           volumetric--deviatoric
           stress increments with an equal stress
           magnitude of
           1~Pa.\label{fig:rendulic2}}
\end{figure}

Figure~\ref{fig:rendulic2} shows similar results
within the Rendulic plane but at a more advanced stage of
loading.
At the reference strain $\varepsilon_{11}=-0.9\%$ the material is
still hardening, but the stress has advanced to 78\% of the peak strength
(see Fig.~\ref{fig:StressStrain}).
The strain contributions in Fig.~\ref{fig:rendulic2}a are
for total strain
increments of equal magnitude
($2\times 10^{-6}$, as in Fig.~\ref{fig:rendulic}b),
and the dilation tendency of the irreversible strain is more
pronounced than at the lower
strain (as in Fig.~\ref{fig:rendulic}c),
with an upward slope toward the left.
The response envelopes in Fig.~\ref{fig:rendulic2}b are for strain increments
that produce equal-magnitude increments of stress.
These results can be compared with those in Fig.~\ref{fig:rendulic}d
for the earlier strain,
as the axes of the two figures share the same aspect ratio.
The larger dilation tendency of irreversible strain is clearly expressed
in Fig.~\ref{fig:rendulic2}.
The regular, elliptic shape of the reversible strain
suggests a linear relationship between the tensors of incremental
strain and stress,
and the more pinched shape of the ellipse
indicates a greater anisotropy of the reversible moduli.
\par
\begin{figure}
  \centering
  \includegraphics{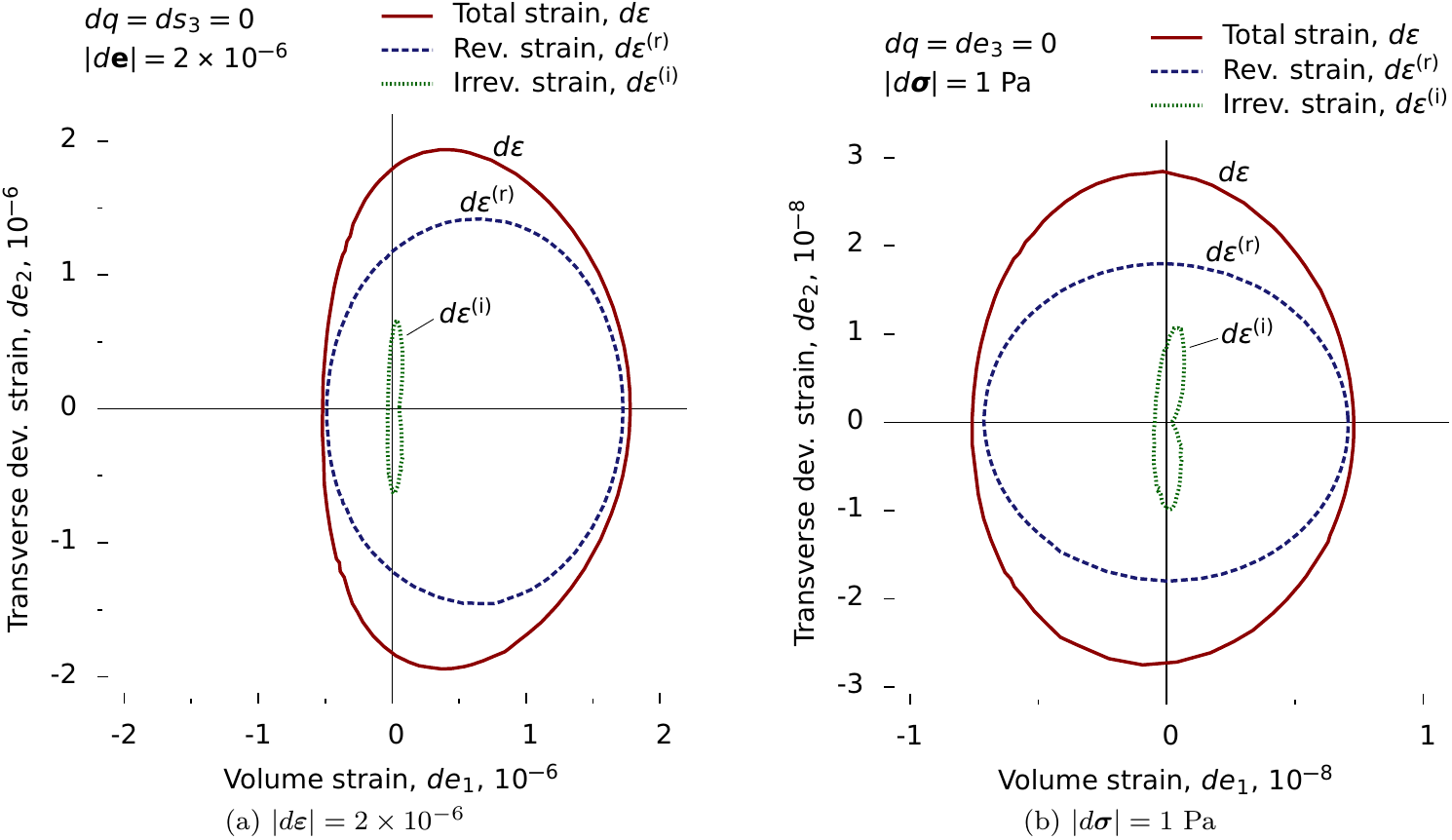}
%
  \caption{Strains for stress probes
           in the plane of volumetric
           and transverse-deviatoric stress
           (transverse-orthogonal planes),
           conducted at the strain
           $\varepsilon_{11}=-0.3\%$:
           (a) loci of volumetric and transverse-deviatoric
           strains for total
           strain increments 
           of equal magnitude; and
           (b) response envelopes of
           strain increments that resulted from
           volumetric and transverse-deviatoric
           stress increments with an equal stress
           magnitude of 1~Pa.
           \label{fig:s1s2}}
\end{figure}

The transverse-orthogonal planes
in Fig.~\ref{fig:s1s2} are of probes in which
stress increments were confined to the
$\vec{\mathbf{s}}_{1}$--$\vec{\mathbf{s}}_{2}$
plane of volumetric
and transverse-deviatoric directions,
a type of result not yet reported in the literature.
As with the previous Figs.~\ref{fig:piplane} and~\ref{fig:rendulic},
this figure is for an initial probe state of
$\varepsilon_{11}=-0.3\%$,
at which the material had attained 55\% of its peak strength.
The magnitudes of the
total strain increments were $2\times 10^{-6}$ for the
results in Fig.~\ref{fig:s1s2}a; whereas, the magnitudes
of the stress increments were 1~Pa for the
response envelopes in
Fig.~\ref{fig:s1s2}b.
Unlike previous figures, the locus of total strain
in Fig.~\ref{fig:s1s2}a is not a circle, since loading
was confined to \emph{stress} within the
$\vec{\mathbf{s}}_{1}$--$\vec{\mathbf{s}}_{2}$ plane,
and some out-of-plane (deviatoric) strains accompanied
the in-plane strains of the figure
(contrarily,
with Figs.~\ref{fig:piplane}b, \ref{fig:rendulic}b,
and \ref{fig:rendulic2}a,
\emph{strains} were confined to their
respective planes).
The most notable feature of this figure
is the significant irreversible strain that occurs in
the \emph{transverse-deviatoric direction}.
In conventional plasticity with a single yield surface,
the surface should
pass through the $\vec{\boldsymbol{\mathbf{s}}}_{2}$
vertical axes of Fig.~\ref{fig:s1s2}
for the axisymmetric conditions produced by
the initial triaxial loading, and
instead of a loop, the irreversible response
$d\mathbf{e}^{\text{(i)}}$ should be a single point
at the origin of these plots.
The vertical transverse-deviatoric irreversible
strains in this figure
are \emph{tangent} to (and lie within)
the conventional yield surface,
and the occurrence of non-zero strains
in this tangent direction is one of several deviations
from conventional plasticity that are addressed
in Section~\ref{sec:irreversible}.
As with the other series of probes,
the reversible response envelope in
Fig.~\ref{fig:s1s2}b is an almost perfect ellipse.
\section{\large Analysis of reversible strains}\label{sec:reversible}
We now separately analyze the reversible and irreversible strain
increments
from at least 240 probes at each of several strains
during the initial phase of constant-$p$ triaxial compression.
The strain increments are analyzed in the context of conventional
elasto-plasticity, with
the reversible strains being analyzed in the current section.
The elliptic shape of the reversible response envelope in
Fig.~\ref{fig:piplane}c 
is evidence of anisotropy in the reversible stiffness,
as the ellipse of reversible strain would appear as a circle for an isotropic
material.
Many of the studies listed
in Tables~\ref{table:studies}--\ref{table:2Dprinciples}
had also found the elastic (or reversible) response envelope to be
modestly elliptic
\cite{Bardet:1994a,AlonsoMarroquin:2005b,Calvetti:2003a,Plassiard:2009a,Sibille:2007a}.
Because our assembly began with an isotropic stress and an
isotropic fabric
before undergoing the initial monotonic axisymmetric
triaxial compression,
we would expect
transverse isotropy to be induced in the stiffness.
If so, the reversible strain $d\boldsymbol{\varepsilon}^{\text{(r)}}$
should have a linear incremental form, with reversible
compliance operator
$\boldsymbol{\mathcal{C}}^{\text{(r)}}$, as
\begin{equation}\label{eq:compliances}
  \renewcommand{\arraystretch}{1.1}
  [d\boldsymbol{\varepsilon}^{\text{(r)}}] =
    \boldsymbol{\mathcal{C}}^{\text{(r)}}(d\boldsymbol{\sigma}),\quad
  \begin{bmatrix}
    d\varepsilon_{1}^{\text{(r)}} \\
    d\varepsilon_{2}^{\text{(r)}} \\
    d\varepsilon_{3}^{\text{(r)}}
  \end{bmatrix}
  =
  \begin{bmatrix}
    \bar{\mathbf{C}}
  \end{bmatrix}
  \begin{bmatrix}
    d\sigma_{1} \\
    d\sigma_{2} \\
    d\sigma_{3}
  \end{bmatrix},\quad
  \begin{bmatrix}
    \bar{\mathbf{C}}
  \end{bmatrix}
  =
  \begin{bmatrix}
    1/E_{1} & -\nu_{21}/E_{2} & -\nu_{21}/E_{2} \\
    -\nu_{12}/E_{1} & 1/E_{2} & -\nu_{2}/E_{2} \\
    -\nu_{12}/E_{1} & -\nu_{2}/E_{2} & 1/E_{2}
  \end{bmatrix}
\end{equation}
%
and as a further condition of transverse isotropy,
$\nu_{21}/E_{2}=\nu_{12}/E_{1}$, leaving
four independent parameters:
the two moduli, $E_{1}$ and $E_{2}$, and the two
Poisson ratios, $\nu_{21}$ and $\nu_{2}$.
\par
We computed best-fit values of the reversible
compliance $[\bar{\mathbf{C}}]$
for a range of initial probe states: from the initial
isotropic assembly ($\varepsilon_{11}=0$)
to the strain at peak strength ($\varepsilon_{11}=-5.3\%$,
Fig.~\ref{fig:StressStrain}).
At each strain,
we used the results of over two hundred stress-controlled probes
within each of the three orthogonal planes,
$\vec{\mathbf{e}}_{1}$--$\vec{\mathbf{e}}_{2}$,
$\vec{\mathbf{e}}_{1}$--$\vec{\mathbf{e}}_{3}$,
and $\vec{\mathbf{e}}_{2}$--$\vec{\mathbf{e}}_{3}$.
The best-fit was a projection of the simulation data~---
the generalized increments $[d\mathbf{s}]$ and $[d\mathbf{e}]$~---
onto the nine-dimensional space of the components of
the generalize compliance matrix $[\mathbf{C}]$, such that
$[d\mathbf{e}] \approx [\mathbf{C}][d\mathbf{s}]$.
The Cartesian components of $[\bar{\mathbf{C}}]$ were then
computed as
$[\bar{\mathbf{C}}]=[\mathbf{E}][\mathbf{C}][\mathbf{E}]^{\mathsf{T}}$.
\par
\begin{figure}
  \centering
  \includegraphics{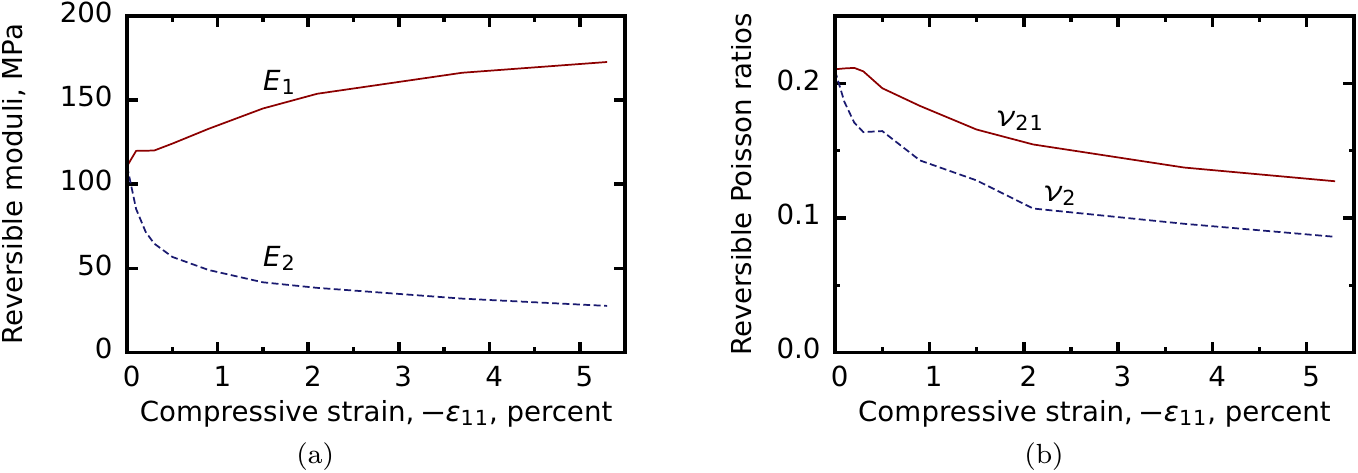}
  \caption{Reversible moduli and Poisson ratios
           for constant-$p$ triaxial compression.
           \label{fig:moduli}}
\end{figure}

Figure~\ref{fig:moduli}
summarizes the reversible moduli and Poisson ratios
at eight strains during constant-$p$ triaxial compression.
Starting from an isotropic condition, anisotropy
was soon expressed at the smallest strain for which the
compliances were computed ($\varepsilon_{11}=-0.1\%$).
The axial
modulus $E_{1}$ increased modestly
from values of 111~MPa to 173~MPa across the range
of strains.
The greatest change, however, was in the lateral modulus
$E_{2}$, which severely
degraded from 111~MPa to 28~MPa.
As such,
the ratio of stiffnesses, $E_{1}/E_{2}$, increased throughout
loading, reaching a ratio greater than 6
at the peak strength.
The Poisson ratios $\nu_{21}$ and $\nu_{2}$ were reduced
across the range of strains (Fig.~\ref{fig:moduli}b).
\par
These results can be interpreted in the context
of the volumetric effects that are typically observed
during monotonic triaxial loading of sands.
The three most common laboratory conditions
are triaxial compression with constant lateral
stress ($\varepsilon_{11}<0$,
$d\sigma_{22}=d\sigma_{33}=\text{constant}$),
constant-$p$ triaxial compression
($\varepsilon_{11}<0$,
$p=\text{constant}$,
$\varepsilon_{22}=\varepsilon_{33}$), and
undrained (isochoric) triaxial compression
($\varepsilon_{11}<0$, $dv=de_{3}=0$,
$\varepsilon_{22}=\varepsilon_{33}$).
\begin{enumerate}
\item
With the first type of laboratory procedure,
the volume change of sand is typically negative (contractive)
at the start of loading but can transition to positive
(dilative) during the course of loading,
and is most dilative near the peak strength.
The stiffness anisotropy that was induced by the
triaxial loading of our simulations
(Fig.~\ref{fig:moduli}) is consistent
with these trends, as the ratio of \emph{reversible}
volume change,
$dv^{\text{(r)}}/(-\varepsilon_{11}^{\text{(r)}})=2\nu_{21}E_{1}/E_{2}-1$,
was $-0.58$ (contractive) at the start of loading,
transitioned to a neutral rate of zero
at a strain of about 0.9\%,
and was 0.59 (dilative) at the strain of 5.3\%,
at the peak strength.
\item
The results of our constant-$p$ simulations are
similar to those of laboratory tests on medium dense sands:
the measured rate of volume change was zero at the start of
loading, was negative for strains between 0 and 0.2\%,
but became dilative at larger strains
and was strongly positive at the peak strength.
The reversible moduli are consistent
with these results,
with a reversible rate of volume change
$dv^{\text{(r)}}/(-\varepsilon_{11}^{\text{(r)}})=
3(1-\nu_{2}+\nu_{21}-E_{2}/E_{1})/(1-\nu_{2}+\nu_{21}+2E_{2}/E_{1})$
that was zero at the start of loading
and become strongly dilative ($=3.1$) at the peak strength.
\item
With laboratory undrained (isochoric)
triaxial compression tests of sands,
the mean stress typically changes as the specimen is
being loaded.
Except for very densely packed specimens,
the mean stress typically decreases at the start of loading
and can either continue to decrease (contractive behavior) or begin
to increase (dilative behavior) at larger strain.
For the reversible moduli in Fig.~\ref{fig:moduli},
the rate of change of the mean stress
$dp/(-d\varepsilon_{11})$ would be negative at the start of
loading, but would transition to a positive rate at a strain
of less than 0.1\% and then continue to increase during loading.
\end{enumerate}
These trends in reversible volumetric behavior result from
the anisotropies of stiffness that were induced by the
initial monotonic loading, and
for each type of laboratory test,
a contractive or dilative tendency can be attributed,
at least partially, to
the induced anisotropy of the reversible stiffness.
These trends, of course, are of the \emph{reversible rates},
the ratio $dv^{\text{(r)}}/(-d\varepsilon_{11}^{\text{(r)}})$,
computed from the reversible moduli and Poisson ratios,
rather than a total rate $dv/(-d\varepsilon_{11})$
that includes both reversible and irreversible strains.
The irreversible strains are analyzed
in the next section.
\par
\begin{figure}
  \centering
  \includegraphics{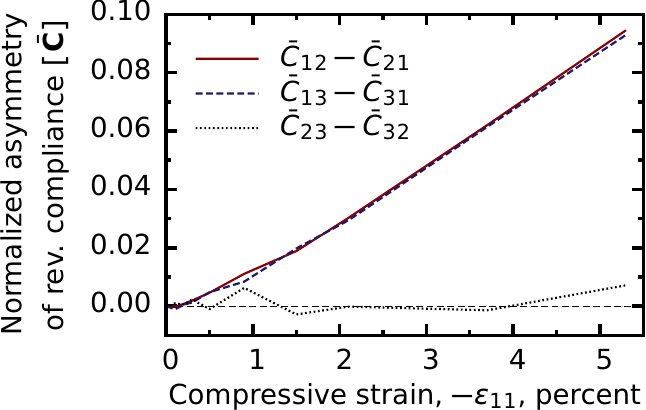}
  \caption{Asymmetry of the reversible compliance matrix $[\bar{\mathbf{C}}]$,
            expressed in the differences of its off-diagonal terms.
            The differences are normalized by dividing by the average
            values of the same off-diagonal terms.
            \label{fig:assymetry}}
\end{figure}

To complete the analysis of reversible behavior,
we considered the possible asymmetry of the reversible
compliance matrix $[\bar{\mathbf{C}}]$.
The Poisson ratios that are plotted in in Fig.~\ref{fig:moduli}b
were computed from the averages of the off-diagonal terms
of $[\bar{\mathbf{C}}]$ in Eq.~(\ref{eq:compliances}).
To test the symmetry of $[\bar{\mathbf{C}}]$,
we computed the off-diagonal differences
$\bar{C}_{12} - \bar{C}_{21}$,
$\bar{C}_{13} - \bar{C}_{31}$, and
$\bar{C}_{23} - \bar{C}_{32}$
(recall that all nine components of $[\bar{\mathbf{C}}]$ were
found as a best-fit among over two hundred probes).
The results are shown in Fig.~\ref{fig:assymetry}, in which 
each difference has been normalized by dividing it by
the corresponding off-diagonal average
(for example, the difference $\bar{C}_{12} - \bar{C}_{21}$ has been
divided by $\frac{1}{2}(\bar{C}_{12} + \bar{C}_{21})$).
The results clearly reveal a significant asymmetry of
the compliance matrix: the compliances $\bar{C}_{12}$ and $\bar{C}_{13}$
unmistakably differ from their counterparts, and the difference
grows with increasing strain.
Taking the na\:{i}ve approach of computing all nine compliances
does result in some small inconsistencies,
with small erratic differences $\bar{C}_{23}-\bar{C}_{32}$
and tiny differences between $\bar{C}_{12}$ and $\bar{C}_{13}$ that
are at variance with the presumed transverse isotropy.
These inconsistencies are small and inconstant, however,
and they are certainly
insignificant when compared with the persistent asymmetry
of the reversible compliance that grows with increasing strain.
This result demonstrates that the reversible operator
$\boldsymbol{\mathcal{C}}^{\text{(r)}}$ can not be strictly derived
from a strain energy density function.
\section{\large Analysis of irreversible strains}\label{sec:irreversible}
We now consider the incremental irreversible behavior in relation to
conventional elasto-plasticity and
evaluate the consistency with (or divergence from) this constitutive
framework.
This analysis reveals new complexities in the
irreversible response of granular materials.
With conventional elasto-plasticity,
the strain increment 
is the sum of reversible and irreversible parts, and
in the previous section, we found that the
reversible part, $d\boldsymbol{\varepsilon}^{\text{(r)}}$
or $d\mathbf{e}^{\text{(r)}}$,
was a linear function of the stress increment:
a compliance operator $\mathbf{C}$ multiplied by the stress increment.
Because the particles' contact mechanism is rate-independent,
the bulk material behavior is also rate-independent and the
functional relationship between the irreversible strain and
the stress increment,
$d\boldsymbol{\varepsilon}^{\text{(i)}}=\boldsymbol{\mathcal{C}}^{\text{(i)}}(d\boldsymbol{\sigma})$,
must be homogeneous of degree~1
with respect to the stress increment $d\boldsymbol{\sigma}$.
The
rate-independent constitutive function
$\boldsymbol{\mathcal{C}}^{\text{(i)}}$
can depend upon the direction of the stress increment,
$d\boldsymbol{\sigma}/|d\boldsymbol{\sigma}|$, as well as on
its magnitude $|d\boldsymbol{\sigma}|$
(see \cite{Hill:1959a} and the proposition of tensorial zones
in \cite{Darve:1990a}).
In conventional elasto-plasticity,
the relationship $\boldsymbol{\mathcal{C}}^{\text{(i)}}$
is assumed
to be incrementally bi-linear with two tensorial zones that are
half-spaces
separated by a hyper-plane (yield surface)
having the unit normal $\mathbf{f}$.
The irreversible
strain increment of conventional elasto-plasticity is expressed as
\begin{equation}\label{eq:Classical}
  [d\mathbf{e}^{\text{(i)}}] =
  \begin{cases}
    0, & [\mathbf{f}]^{\mathsf{T}}[d\mathbf{s}] \le 0 \\
    \displaystyle
    \frac{1}{h}[\mathbf{g}][\mathbf{f}]^{\mathsf{T}}[d\mathbf{s}], &
      [\mathbf{f}]^{\mathsf{T}}[d\mathbf{s}] > 0
  \end{cases}
  \qquad\text{or}\qquad
  [d\boldsymbol{\varepsilon}^{\text{(i)}}] =
  \begin{cases}
    0, & [\bar{\mathbf{f}}]^{\mathsf{T}}[d\boldsymbol{\sigma}] \le 0 \\
    \displaystyle
    \frac{1}{\bar{h}}[\bar{\mathbf{g}}][\bar{\mathbf{f}}]^{\mathsf{T}}[d\boldsymbol{\sigma}], &
      [\bar{\mathbf{f}}]^{\mathsf{T}}[d\boldsymbol{\sigma}] > 0
  \end{cases}
\end{equation}
for the alternative systems of generalized stress and strain vectors,
$[d\mathbf{s}]$ and $[d\mathbf{e}^{\text{(i)}}]$,
and of their Cartesian counterparts
$[d\boldsymbol{\sigma}]$ and $[d\boldsymbol{\varepsilon}^{\text{(i)}}]$
(see Eqs.~\ref{eq:Esystem}--\ref{eq:Ssystem}).
In these expressions, $[\mathbf{f}]$ is the unit vector of the yield
direction; $[\mathbf{g}]$ is the unit flow direction,
which is the direction
of the irreversible strain $[d\mathbf{e}^{\text{(i)}}]$;
and $h$ is the scalar plastic modulus.
Over-bars are used to distinguish the Cartesian counterparts, noting
that the two systems~--- generalized and Cartesian~--- are formed from triads
of orthonormal basis vectors (Eq.~\ref{eq:Esystem}), and
one can readily shift between the two systems: with
$[\bar{\mathbf{f}}]=[\mathbf{f}][\mathbf{E}]^{\mathsf{T}}$,
$[\bar{\mathbf{g}}]=[\mathbf{E}][\mathbf{g}]$,
and $h=\bar{h}$.
\par
We now investigate this question: does the conventional plasticity of
Eqs.~(\ref{eq:Classical})
match the incremental behavior of the simulations?
The question has four aspects,
which correspond to the principles~3--6 of the
Introduction:
(3)~whether a stress direction of purely elastic response
exists;
(4)~whether a single yield surface
with normal direction $\mathbf{f}$ separates two tensorial
half-spaces of
irreversible and reversible behaviors;
(5)~whether irreversible strain is uniformly
in a single direction $\mathbf{g}$
that is independent of the stress increment $d\boldsymbol{\sigma}$;
and (6)~whether the irreversible strain
$d\mathbf{e}^{\text{(i)}}$ is proportional to
the projected stress $[\mathbf{f}]^{\mathsf{T}}[d\boldsymbol{\sigma}]$.
Based upon the
discussion of Fig.~\ref{fig:piplane},
one must suspect that the first
question is answered in the negative, since we had found that
irreversible deformation
occurs during both
compressive loading and extensional unloading,
which suggests that a region of purely elastic behavior, if it exists at all,
is smaller than the strain increment of 2$\times 10^{-6}$ that was used in our
simulations.
Answers to the four questions, although somewhat hidden within
Figs.~\ref{fig:piplane}--\ref{fig:s1s2}, are clearly revealed
in special polar plots of the simulation results.
\par
\begin{figure}
  \centering
  \includegraphics{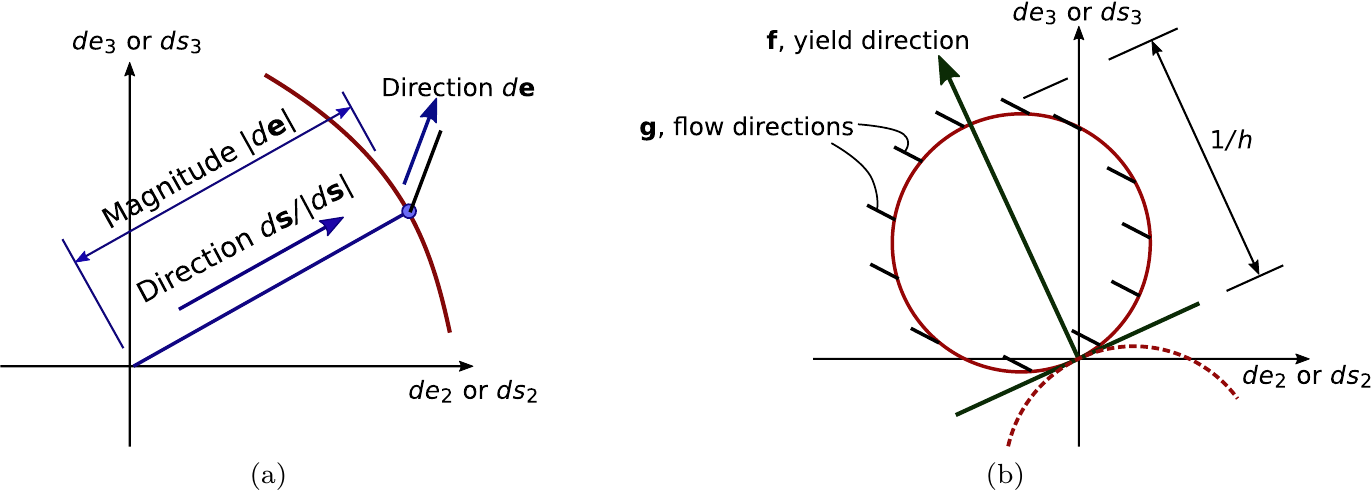}
%
  \caption{Plotting the irreversible stress--strain
           relationship with polar plots and spars
           (as in Eq.~\ref{eq:Sphere1}).
           The example shown is for behavior
           within the $\vec{\mathbf{e}}_{2}$--$\vec{\mathbf{e}}_{3}$
           plane:
           (a)~points represent magnitudes and directions
           of the irreversible strain; and
           (b)~conformity with conventional elasto-plasticity
           creates a circular polar locus.
           \label{fig:plots}}
\end{figure}
\begin{figure}
  \includegraphics{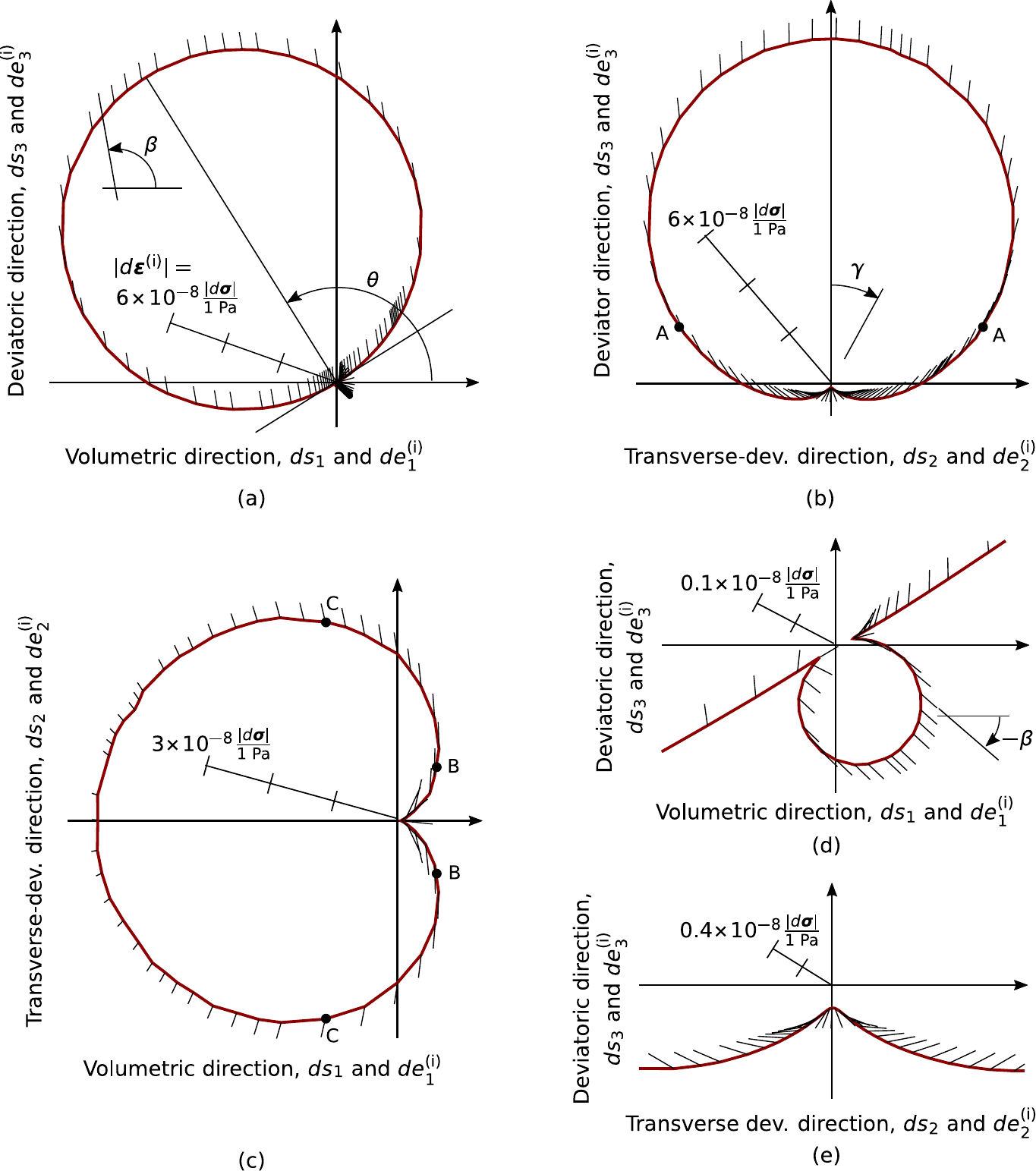} 
  \caption{Polar plots showing the results of multiple probes in the
           the generalized space of irreversible
           strain $d\vec{\mathbf{e}}^{\text{(i)}}$
           and of stress $d\vec{\mathbf{s}}$.
           As with Eq.~(\ref{eq:Sphere1}),
           radial distance from the origin represents
           the irreversible strain
           magnitude $|d\boldsymbol{\varepsilon}^{\text{(i)}}|$
           produced by a
           stress increment of the
           corresponding direction $d\mathbf{s}/|d\mathbf{s}|$.
           Short spars show the directions of the irreversible strains:
           (a)~Rendulic plane of volumetric and deviatoric increments,
           (b)~pi-plane of deviatoric and transverse-deviatoric increments,
           (c)~plane of volumetric and transverse-deviatoric increments,
           (d)~detail of the Rendulic plane, and
           (e)~detail of the pi-plane. 
           The probes are
           at a reference strain $\varepsilon_{11}=-0.9\%$
           and were conducted with Cattaneo--Mindlin contacts.
           \label{fig:circles}}
\end{figure}

Our plotting method is illustrated in Fig.~\ref{fig:plots}
and is applied in
Fig.~\ref{fig:circles},
the latter showing results of the material's response
in three series of probes,
with each probe sharing a common
magnitude of the stress increment, $|d\mathbf{s}|$.
Figure~\ref{fig:circles} is for probes initiated
at the compressive strain $\varepsilon_{11}=0.9\%$,
at which the deviatoric stress had advanced to 78\% of the peak strength.
The three series of probes were conducted in the three orthogonal
planes: the Rendulic plane $d\vec{\mathbf{e}}_{1}$--$d\vec{\mathbf{e}}_{3}$,
the octahedral pi-plane $d\vec{\mathbf{e}}_{2}$--$d\vec{\mathbf{e}}_{3}$,
and the plane $d\vec{\mathbf{e}}_{1}$--$d\vec{\mathbf{e}}_{2}$
of volumetric and transverse-deviatoric increments.
Because generalized stress and strain, $[d\mathbf{s}]$ and
$[d\mathbf{e}]$, share the same basis vectors, the planes in
Fig.~\ref{fig:circles} represent both stress and strain.
Each sub-figure depicts two different aspects of behavior,
as illustrated in Fig.~\ref{fig:plots}a.
The closed solid (red) line is a polar plot of
the magnitude of the strain increment $|d\mathbf{e}|$,
and it
is directly related to the yield condition $\mathbf{f}$.
The thin short spars show flow directions $\mathbf{g}$,
and these spars are discussed further below.
Each closed solid (red) line is the locus of points
\begin{equation}\label{eq:Sphere1}
  \frac{|d\mathbf{e}^{\text{(i)}}|}{|d\mathbf{s}|}
  \frac{d\mathbf{s}}{|d\mathbf{s}|}
  \;\longrightarrow\;
  \text{Figure }\ref{fig:circles}\text{ locus}
\end{equation}
of the over eighty probes for one of the three series of probes.
That is, the radial distance from the origin of a plot
to a point on the solid curve is the magnitude of the
irreversible strain increment $|d\mathbf{e}^{\text{(i)}}|$ that results from a
stress increment of magnitude $|d\mathbf{s}|$
(i.e. the first quotient in Eq.~\ref{eq:Sphere1}).
The radial line is oriented in
the \emph{direction} of the stress increment, $d\mathbf{s}/|d\mathbf{s}|$,
that produced the strain increment
(this quotient is a unit vector in stress space, whereas the
radial distance $|d\mathbf{e}^{\text{(i)}}|$
is a scalar measure of the irreversible strain).
\par
We now return to the conventional plasticity
of Eq.~(\ref{eq:Classical}):
if the irreversible strain conforms to this equation,
then the locus of points in Eq.~(\ref{eq:Sphere1}) will be a sphere
(Fig.~\ref{fig:plots}b, noting that the lower sphere is disallowed
by the condition $[\mathbf{f}]^{\mathsf{T}}[d\mathbf{s}] > 0$
in Eq.~\ref{eq:Sphere1}).
That is,
if the conventional plasticity of Eq.~(\ref{eq:Classical}) holds,
then the solid curves in the cross-sections of
Fig.~\ref{fig:circles} should be circles,
regardless of the plane that
is being represented in the generalized space of $d\mathbf{s}^{\text{(i)}}$.
The generating sphere
of these circles would pass through the origin, its radius would be
$1/h$, and the tangent plane of the sphere at the origin
would have the unit normal direction $\mathbf{f}$.
These characteristics for a plot of 
Eq.~(\ref{eq:Classical})
are independent of the flow direction $\mathbf{g}$.
As such,
we can test whether the material conforms to the ideal yield condition
by observing the shapes
of the solid curves in each of the planes of Fig.~\ref{fig:circles}.
Note that similar results would also apply to the space of Cartesian
increments, $[d\boldsymbol{\varepsilon}^{\text{(i)}}]$
and $[d\boldsymbol{\sigma}]$, as this system also has
an orthonormal basis.
%
\par
The plots in Figs.~\ref{fig:plots}
and~\ref{fig:circles} also include a series
of short thin spar lines that show the \emph{directions}
of the irreversible
strain $d\mathbf{e}^{\text{(i)}}$
for different directions of the stress increment,
$d\mathbf{s}/|d\mathbf{s}|$,
with the direction of $d\mathbf{e}^{\text{(i)}}$
projected onto the plane of the plot.
That is,
the radial direction from the origin to a point on the closed curved
is the stress direction $d\mathbf{s}/|d\mathbf{s}|$,
the radial distance is the strain magnitude
$|d\mathbf{e}|/|d\mathbf{s}|$,
and the spar shows the direction of the irreversible strain
$[d\mathbf{e}^{\text{(i)}}]/|d\mathbf{e}^{\text{(i)}}|$
that results from the stress increment
(Fig.~\ref{fig:plots}a, where the strain
direction vector is projected onto each particular
plane).
Conformity with conventional elasto-plasticity
(Eq.~\ref{eq:Classical} and principle~5) demands a uniform
direction and length of these spars.
\par
Figure~\ref{fig:circles}a shows these results
for the Rendulic plane of volumetric
and deviatoric stress and strain.
The radial ruler in this plot gives the scale of
strain magnitude, $|d\mathbf{e}|$ or $|d\boldsymbol{\varepsilon}|$,
for the closed curve.
The closed curve is an almost perfect circle,
meaning that the conventional plasticity of Eq.~(\ref{eq:Classical})
does seem to
apply for axisymmetric \emph{triaxial} probes in which various combinations
of mean stress $dp$ and deviator stress $dq$ are applied
(i.e., probes in the Rendulic plane
$d\vec{\mathbf{s}}_{1}$--$d\vec{\mathbf{s}}_{3}$).
The diameter of the circle in Fig.~\ref{fig:circles}a is the
inverse plastic modulus, such that $h=17.6$~MPa.
The circle is tilted at angle $\theta=121^{\circ}$ from
the $d\vec{\mathbf{s}}_{1}$ axis,
with the unit yield vector $\mathbf{f}$ would have direction
$[\cos\theta, 0, \sin\theta]^{\mathsf{T}}$ in the generalized stress-space
$[d\mathbf{s}]$.
This yield direction
considers only behavior within the Rendulic plane,
whereas the out-of-plane
behavior is considered below.
The yield direction $\mathbf{f}$
corresponds to 
a yield ``friction angle''
$\sin^{-1}((d\sigma_{11}-d\sigma_{33})/(d\sigma_{11}+d\sigma_{33}) = 33^{\circ}$
for these triaxial conditions.
At the stage of loading of the figure,
the mobilized friction angle
$\sin^{-1}((\sigma_{11}-\sigma_{33})/(\sigma_{11}+\sigma_{33})$
was $30^{\circ}$, somewhat less than the yield angle.
\par
Upon close inspection, however,
a second but much smaller circle is also present in
the Rendulic plane,
below and to the right of the larger circle.
A detail of
this circle is shown in Fig.~\ref{fig:circles}d,
demonstrating that a small amount of dissipative, irreversible
strain occurs during
a reversal of triaxial compression.
The two circles, forward and reverse (loading and unloading),
are nearly (but not quite)
tangent to each other and to the same oblique line in the
Rendulic plane that defines the perpendicular angle $\theta$.
As with the generalized plasticity model of
Zienkiewicz and Pastor \cite{Pastor:1986a},
irreversible
deformation is activated for opposite directions
of the stress increment~--- for both loading and unloading~---
with yield directions
$\mathbf{f}_{\text{loading}}=-\mathbf{f}_{\text{unloading}}$.
In regard to the conventional plasticity of
Eq.~(\ref{eq:Classical}), the irreversible
strain increment for the first, ``elastic'' case is not zero,
but instead has a form
similar to that of the second, ``plastic'' case, but with a much larger modulus
$h$ (the plastic moduli for loading and unloading are
17.6~MPa and 720~MPa, respectively).
In Fig.~\ref{fig:circles}d, we also note that the two circles,
large and small, do not exactly meet at a common point, 
demonstrating that irreversible strain occurs in all
360$^{\circ}$ of loading within the Rendulic plane,
thus abrogating principle~3 of a finite elastic domain
within the axisymmetric Rendulic plane.
This result, which can only be discerned by applying very
small strain increments, was not reported in past studies.
The absence of a loading direction that produces purely reversible
strain is shown also to apply in the two other planes of loading.
\par
Returning to the Rendulic plane of
Fig.~\ref{fig:circles}a and ignoring for the moment
the small unloading circle at the bottom right
of this figure, the spars are almost uniformly aligned in
the same direction around the larger circle,
regardless of position on the circle
(i.e., regardless of the direction of the stress increment).
Although an important exception will be noted below,
this result means that a single flow direction $\mathbf{g}$ applies
to axisymmetric triaxial loading within the Rendulic plane
(see Eqs.~\ref{eq:Classical}).
A similar uniformity
of flow direction in the Rendulic plane
of axisymmetric loading has also been found in
2D simulations \cite{Bardet:1994a},
in 3D simulations \cite{Kishino:2003a,Plassiard:2009a,Froiio:2010a},
and in soil experiments
\cite{Lewin:1970a,Tatsuoka:1974b,Anandarajah:1995a,Darve:2005b}.
The conclusion of a nearly uniform direction in the
Rendulic plane would also be suspected from the results in
Figs.~\ref{fig:rendulic}d and~\ref{fig:rendulic2}b:
the curves of irreversible strain $d\mathbf{e}^{\text{(i)}}$
appear as straight lines in these views, confirming
a nearly uniform direction of $d\mathbf{e}^{\text{(i)}}$ in the
$d\vec{\mathbf{e}}_{1}$--$d\vec{\mathbf{e}}_{3}$ plane.
This observation is somewhat deceptive, however, as a full plane
of fanned flow directions might appear as a line in the edge-wise perspective
of these figures, as will be demonstrated below.
\par
The flow spars around the larger circle
in Fig.~\ref{fig:circles}a are oriented upward and to the
left, at
an angle of $\beta = 100^{\circ}$ relative to the $d\vec{\mathbf{e}}_{1}$
direction of purely isotropic compression,
indicating that the incremental irreversible strain is dilative.
This orientation corresponds to a dilatancy angle
$\psi^{\text{(i)}}=\sin^{-1}((d\varepsilon_{11}^{\text{(i)}}+d\varepsilon_{33}^{\text{(i)}})/(d\varepsilon_{11}^{\text{(i)}}-d\varepsilon_{33}^{\text{(i)}}))$
of 10$^{\circ}$ and an irreversible dilation
rate $dv^{\text{(i)}}/|d\varepsilon_{11}^{\text{(i)}}|$
of 0.43. 
As with most studies listed in Table~\ref{table:studies}, the orientations
of yield and flow, $\mathbf{f}$ and $\mathbf{g}$, do not coincide,
with $\beta\neq\theta$,
and the irreversible response is clearly non-associative.
\par
For the small circle that corresponds to triaxial unloading
(Fig.~\ref{fig:circles}d),
the irreversible strain direction is almost
uniformly downward
and to the right with an angle $\beta=-45^{\circ}$,
such that unloading increments produce
irreversible volume compression rather
than dilation.
Figure~\ref{fig:circles}d shows that irreversible
strain occurs in two different directions for triaxial
(Rendulic) loading and unloading,
with
$\mathbf{g}_{\text{loading}}\neq\mathbf{g}_{\text{unloading}}$.
Similarly,
the results of Lewin and Burland \cite{Lewin:1970a}
and Tatsuoka and Ishihara \cite{Tatsuoka:1974b}
also indicated a difference in the directions of plastic
strain for stress increments that produce loading
and those that produce unloading,
a difference that is accommodated with generalized
plasticity \cite{Zienkiewicz:1984a,Pastor:1986a}.
\par
A seemingly abrupt,
discontinuous change in the direction of the irreversible
strain (the directions of the spars) occurs at the two cusps in
Fig.~\ref{fig:circles}d, suggesting that the direction of the strain
increment might be indeterminate when the
stress increment is tangential to the yield direction
(directions $\theta=121^{\circ}\pm 90^{\circ}$
in Fig.~\ref{fig:circles}a).
This result is troubling,
as it suggests a violation of Hashiguchi's
continuity requirement \cite{Hashiguchi:1993a},
which holds that the total strain increment
is identical for infinitesimally close loading directions
(e.g., for directions on either side of the cusp
in Fig.~\ref{fig:circles}d).
To evaluate continuity,
we conducted more finely spaced probes at these and at other
cusps, which will be discussed below.
\par
Note, too, that purely volumetric loading~--- increments $d\mathbf{s}$
along the horizontal axis in Fig.~\ref{fig:circles}d~--- produces
irreversible increments of both volumetric and deviatoric strain.
For an increasing pressure (toward the right of the origin
in Fig.~\ref{fig:circles}d), the volume change, although small, is
finite, such that the irreversible reduction of volume $dv^{\text{(i)}}$
is equal to $dp/$2.1~GPa.
This observation is consistent with soil behavior that is described
with ``capped'' yield surfaces that activate plastic deformation
with increasing pressure increments \cite{Lai:2016a}
(note, however, that no such
cap is invoked in the current work).
The volumetric response is much softer for pressure reductions
(toward the left of the origin in Fig.~\ref{fig:circles}d),
with $dv^{\text{(i)}}=dp/$16~MPa.
Cap models typically apply only to volumetric loading (compression),
but the results in Fig.~\ref{fig:circles}d
show that both deviatoric and volumetric irreversible strains
are induced by both volumetric and deviatoric
unloading (downward and to the left in
Fig.~\ref{fig:circles}d),
a result that has not been revealed in previous studies.
\par
We now turn attention to Fig.~\ref{fig:circles}b,
which displays results of the series of isobaric stress probes in the
$d\vec{\mathbf{s}}_{2}$--$d\vec{\mathbf{s}}_{3}$ octahedral pi-plane of
deviatoric loading (i.e., similar to the results displayed
in Fig.~\ref{fig:rendulic2}b).
In a laboratory setting,
such stress increments could be produced with true-triaxial equipment.
The closed solid curve in this figure
was produced in the manner of Eq.~(\ref{eq:Sphere1}).
The curve is clearly not a circle passing through the origin,
meaning that the ideal conventional plasticity
of Eqs.~(\ref{eq:Classical}) does not apply to
true-triaxial deviatoric or to transverse-deviatoric loadings.
For example,
the curve's breadth (in the $ds_{2}$ direction)
is greater than its height above the $ds_{3}$ axis,
indicating softer behavior in the transverse direction.
The curve also passes \emph{below}
the horizontal axis, which means that
deviatoric unloading (with $ds_{3}$ or $dq \le 0$)
produces irreversible strain.
\par
If one assumes that a single yield direction $\mathbf{f}$
applies and that the yield surface is a plane of
orientation $\theta$ and is perpendicular
to the Rendulic plane of Fig.~\ref{fig:circles}a,
then the transverse-deviatoric
$\vec{\mathbf{s}}_{2}$ direction is tangent
to the yield surface.
The fact that irreversible deformation occurs
in these tangential loading directions
($ds_{3}\le 0$ and $de^{\text{(i)}}_{2}\neq0$,
or the 3~o'clock to 9~o'clock directions
in Fig.~\ref{fig:circles}b) 
confirms the long-held conjecture of tangential
plasticity.
Evidence of tangential plasticity has also been revealed
in the 3D simulations of Kishino
\cite{Kishino:2003a} and
Plassiard et al. \cite{Plassiard:2009a}.
The conjecture of tangential plasticity
originated from Hill's analysis of
multi-slip crystal plasticity \cite{Hill:1967a}
and as a possible rationale for softer behavior
in loading directions that could produce certain localized
modes of deformation \cite{Rudnicki:1975a,Hashiguchi:2005a},
and a simple two-mechanism model is investigated in
Section~\ref{sec:multimodel}.
Similar to multi-slip plasticity models, certain
micro-mechanics models of granular materials are
based upon presumed distributions of contact orientations
and upon the normal and tangential interactions on
contact planes of different orientation
\cite{Darve:2005a,Hicher:2007a,Iai:2011a}.
These hypothetical models also lead to tangential plasticity,
although they presume a single, averaged
behavior to all contacts with the same orientation,
assume affine movements among the particles,
and homogenize movements across the space of orientations.
Recent DEM simulations have shown, however,
that micro-scale movements are irregular and non-affine,
with particles seemingly darting and dodging in
a highly irregular manner,
and that contact movements are highly diverse,
even when one considers a subset of contacts having
the same orientation
\cite{Tordesillas:2008a,Kuhn:2016a,Kuhn:2016b}.
Moreover, a fraction of contacts are persistently
non-elastic and continue to undergo frictional slip even
when the direction of bulk deformation is reversed
\cite{Kuhn:2018a}.
\par
The solid line
in Figs.~\ref{fig:circles}b and~\ref{fig:circles}e
is also concave near (and below) the origin
of the figure.
When plotted in the manner of Eq.~(\ref{eq:Sphere1}),
the behavior is clearly incompatible with conventional
elasto-plasticity:
the yield condition (yield surface)
is seemingly cornered; a single yield direction $\mathbf{f}$ does not apply;
and transverse-deviatoric increments $ds_{2}$ and
deviatoric unloading $ds_{3}<0$ 
generate irreversible strains,
a result not revealed in previous experiments or
simulations.
A blunt corner \emph{below the origin} is present in the detail of
Fig.~\ref{fig:circles}e, a result that has not been recognized
in previous experiments and simulations.
In their micro-fissure model for rock,
Rudnicki and Rice \cite{Rudnicki:1975a} showed that a cornered elastic
regime is associated with a reduced stiffness in the
transverse direction, due to
the irreversible strains that would accompany transverse
loading
(note, too, that the \emph{reversible} stiffness is also reduced
in the transverse direction, as seen in Fig.~\ref{fig:moduli}a).
Unlike their micro-fissure model, however, an elastic
regime does not exist in the
$\vec{\mathbf{s}}_{2}$--$\vec{\mathbf{s}}_{3}$ plane,
as the closed curve in Figs.~\ref{fig:circles}b
and~\ref{fig:circles}e fully encloses the origin,
demonstrating that irreversible deformation
occurs in all loading directions within the pi-plane.
\par
Figures~\ref{fig:circles}b and~\ref{fig:circles}e also depict
the \emph{directions} of irreversible strain,
$d\boldsymbol{\varepsilon}^{\text{(i)}}$ or $d\mathbf{e}^{\text{(i)}}$,
as the short spars that emanate from the closed curve.
Each spar begins from a point on the curve that corresponds to the
direction $d\mathbf{s}$ of the applied stress increment.
The results clearly show that irreversible deformation
is not aligned in a uniform direction for different directions
of loading, $[d\mathbf{s}]/|d\mathbf{s}|$.
As one moves from the top of the closed curve, around the sides, and
toward the bottom, the direction of the irreversible strain progressively
points further away from the upward $d\vec{\mathbf{e}}_{3}$ direction.
Therefore, a single flow direction $\mathbf{g}$
(as in Eq.~\ref{eq:Classical})
does not apply to
loading within the deviatoric pi-plane;
instead, the flow direction
is a function of the direction of the stress increment.
This result, which contradicts principle~5, has also
been reported in previous simulations
(see \cite{Kishino:2003a,Tamagnini:2005a,Wan:2014a}).
Moreover, irreversible
strain can occur in any direction within the pi-plane~---
a full 360$^{\circ}$ of directions~---
if the proper stress direction
$d\mathbf{s}/|d\mathbf{s}|$ is applied.
The transverse-deviatoric strain $de_{2}^{\text{(i)}}$ is zero
at the top and bottom of Fig.~\ref{fig:circles}b
(consistent with the initial axial symmetry), but
$de_{2}^{\text{(i)}}$ is largest at points ``A'' rather than
in the $\vec{\mathbf{s}}_{2}$ direction along the horizontal axis.
\par
The spars in Fig.~\ref{fig:circles}b show projections of irreversible
strain onto the isobaric pi-plane, but volume change is also
induced by deviatoric loading.
For the upper two quadrants of the figure (positive deviator stress $q$),
the irreversible volumetric strain $de_{1}^{\text{(i)}}$ is
dilative, but the direction is reversed in the lower half of the diagram,
and the irreversible volume change is compressive for the entire portion
of the pi-plane shown in Fig.~\ref{fig:circles}e.
Such observations are inconsistent with the conventional elasto-plasticity
of Eq.~(\ref{eq:Classical})
and have not been previously reported.
\par
The situation is also complex when we consider loading increments
within the transverse plane $d\vec{\mathbf{e}}_{1}$--$d\vec{\mathbf{e}}_{2}$,
shown in Fig.~\ref{fig:circles}c and unique to the current study.
The horizontal axis of this plot represents
increments of volume (or pressure), with compression to the right, and
the vertical axis corresponds to deviatoric loading,
$ds_{2}$ or $de_{2}$, that is \emph{transverse}
to the axisymmetric triaxial deviator (i.e., transverse to
$ds_{3}$ and $de_{3}$ in Eq.~\ref{eq:Esystem}).
The deviatoric stress increment, $ds_{3}^{\text{(i)}}$ or $dq$,
is zero within this plane.
For loading within this
$d\vec{\mathbf{e}}_{1}$--$d\vec{\mathbf{e}}_{2}$ plane,
the yield condition is bluntly cornered, irreversible flow varies with the
direction of the stress increment $d\mathbf{s}/|d\mathbf{s}|$
(i.e., incremental nonlinearity),
and irreversible strain occurs with all directions of loading.
Dilation occurs for most loading directions,
but compression (contraction) occurs for directions between
the points ``B''.
The transverse-deviatoric strain $de_{2}^{\text{(i)}}$ is largest at
points ``C''.
Irreversible flow, represented by the short spar lines, can occur
in a wide swath of directions~--- a full 360$^{\circ}$ of directions~---
within the plane, depending upon the loading direction
$d\mathbf{s}/|d\mathbf{s}|$.
\par
\begin{figure}
  \centering
  \includegraphics{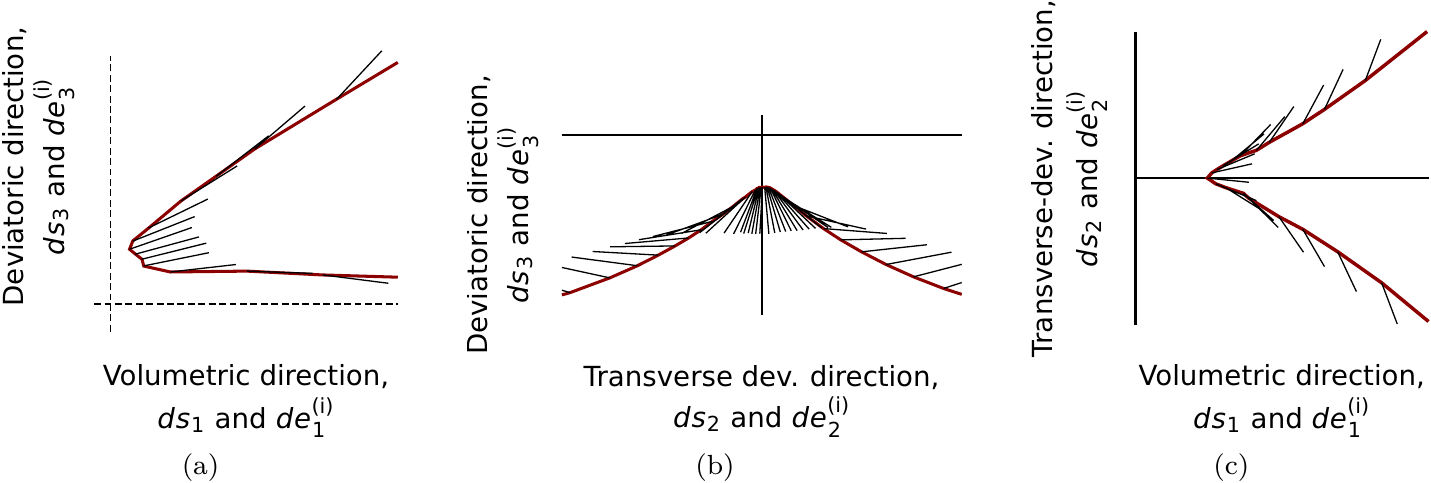}
%
  \caption{Magnified results at cusps:
           (a)~cusp within the Rendulic plane,
           as in Figs.~\ref{fig:circles}a and \ref{fig:circles}d;
           (b)~cusp within the deviatoric plane,
           as in Figs.~\ref{fig:circles}b and \ref{fig:circles}e;
           and
           (c)~cusp within transverse-deviatoric--volumetric
           plane, as in Fig.~\ref{fig:circles}c.
           \label{fig:cusps}}
\end{figure}

Seeming cusps appear in all of the loading planes
of Fig.~\ref{fig:circles}.
In a final set of these special plots,
Fig.~\ref{fig:cusps} magnifies the results at these locations.
In each case, we see that the direction of irreversible
strain (the spars in this figure) does not change abruptly,
which would have indicated a discontinuity in the response
and a violation of the Hashiguchi continuity requirement
described above
\cite{Hashiguchi:1993a}.
The solid lines are not pointed but are rounded, and
as one traces along any of the solid lines in the figure,
the directions of the spars (i.e., the directions
of the irreversible strain) change rapidly
but in a continuous manner.
%
\par
Considering all of the plots in Fig.~\ref{fig:circles},
none of the closed curves showing the irreversible response
passes through the origin of the
$d\vec{\mathbf{e}}_{1}$--$d\vec{\mathbf{e}}_{2}$--$d\vec{\mathbf{e}}_{3}$
system.
In contravention of conventional elasto-plasticity
(principle~3)
and of more general models that are based on neutral
loading directions \cite{Pietruszczak:1986a},
we can conclude that a reversible (elastic) regime, if it exists at all,
is smaller than the strain increments of 2$\times$$10^{-6}$
that were used in these simulations.
This characteristic has been suggested by Royis and Doanh \cite{Royis:1998a}
on the basis of their tests on sands that employed much larger strain
increments.
Alonso-Marroqu\'{i}n and Herrmann \cite{AlonsoMarroquin:2005a} also
arrived at this conclusion
by unloading a DEM assembly of polygons and then reloading on various paths.
Many of our results, however, have not been observed in the previous
studies listed Tables~\ref{table:studies}--\ref{table:3Dprinciples}:
including the irreversible strains that occur during reversed
loading
in the Rendulic plane,
the directions of the transverse-deviatoric (tangential) strain
increments,
the cornered nature of yield within
the pi-plane, etc.
Had we solely relied on Gudehus response envelopes, such as those
in Figs.~\ref{fig:piplane}c
and~\ref{fig:rendulic}d,
these phenomena would have been missed in the current study,
and they were only recognized by plotting data
in a special manner of Figs.~\ref{fig:plots} and~\ref{fig:circles}.
\par
\begin{figure}
  \centering
  \includegraphics{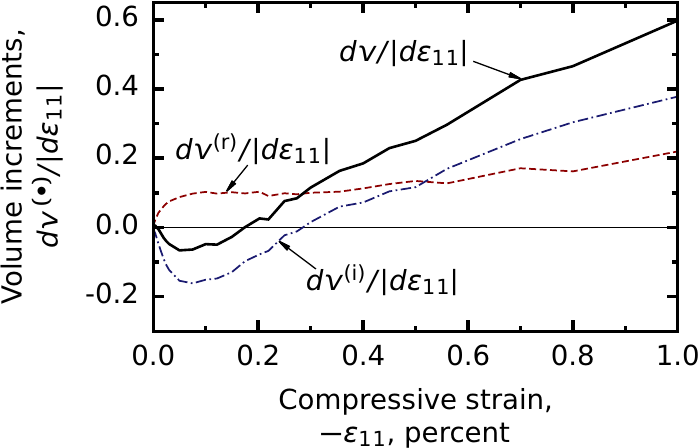}
  \caption{Contributions of reversible and irreversible
           strains to the rate of volume change
           during constant-$p$ triaxial compression.
           \label{fig:volume}}
\end{figure}

As a final observation, Fig.~\ref{fig:volume}
shows the contributions of
the reversible and irreversible strains to the rate of volume
change $dv/|d\varepsilon_{11}|$.
For our loading with initial loading with constant
mean stress (constant-$p$),
the total rate of volume change $dv$
was negative (contractive) at the start of
loading but transitioned to a positive (dilatant)
rate at a strain of 0.2\% (i.e., the ``characteristic state''
of soil behavior).
As concluded in Section~\ref{sec:reversible}, the contribution
of the reversible strain $dv^{\text{(r)}}$ was dilative throughout
the loading, due to the anisotropy that was induced in the stiffness
moduli.
The irreversible volume change $dv^{\text{(i)}}$ transitions
from contractive to dilative and becomes the dominant influence
on volume behavior at strains beyond 0.5\%.
\section{Advanced models}
Among other unusual aspects of our results,
the shapes in
Figs.~\ref{fig:circles}b--\ref{fig:circles}e
demonstrate that behavior is softer in the transverse-deviatoric
direction $d\vec{\mathbf{s}}_{2}$ than would be predicted by
the conventional plasticity of Eq.~(\ref{eq:Classical}).
Several advanced theories of incremental stiffness
have been proposed for modeling such behavior
(see summaries by Bardet \cite{Bardet:1994a} and
Hashiguchi et al. \cite{Hashiguchi:2005a}), and
we considered two modest enhancements of the conventional
elasto-plasticity of Eq.~(\ref{eq:Classical}):
the multi-mechanism model of Mandel
\cite{Hill:1967a,Anandarajah:2008a} and
the tangential plasticity model
of Rudnicki and Rice \cite{Rudnicki:1975a}.
For reasons that are described later,
the most promising approach is a multi-mechanism model, which
we now describe in the context of the simulation results.
\subsection{Multi-mechanism model}\label{sec:multimodel}
The corners in
Figs.~\ref{fig:circles} and~\ref{fig:cusps}, although rounded,
suggest an incrementally multi-linear form of the yield
condition (yield surface), rather than the single-mechanism assumption
of conventional plasticity in Eq.~(\ref{eq:Classical}).
Our simulations were of multi-directional increments
after an initial stage of axisymmetric
triaxial loading.
For this special case of initial loading,
we would expect that the multiple mechanisms
are symmetric with respect to the plane of loading~---
the $d\vec{\mathbf{s}}_{1}$--$d\vec{\mathbf{s}}_{3}$
Rendulic plane~---
and we consider two mechanisms, with yield
directions $\mathbf{f}_{1}$ and $\mathbf{f}_{2}$ that are
mirrored with respect to this plane.
Although it is possible for the two types of yielding
to produce flow $\mathbf{g}$ in the same direction,
the symmetric fanning of
the irreversible spars in Figs.~\ref{fig:circles}b
and~\ref{fig:circles}c suggests that the two mechanisms
produce flow in directions that
are also symmetric
with respect to the
$d\vec{\mathbf{s}}_{1}$--$d\vec{\mathbf{s}}_{3}$ plane.
When both mechanisms are activated, the direction of irreversible
strain can have a component in the transverse
$d\vec{\mathbf{e}}_{2}$ direction, and the magnitude of this transverse
component will depend upon the unbalanced activation of the
two mechanisms.
Moreover,
the large and small circles in
the Rendulic planes of Figs.~\ref{fig:circles}a and~\ref{fig:circles}d
indicate that each of the two mechanisms will produce irreversible
strains for both loading and unloading increments of stress.
In this respect, we diverge from the multi-mechanism model of
Mandel \cite{Mandel:1965a}, who hypothesized distinctly reversible
and irreversible tensorial zones with a specific
flow direction for each mechanism.
\par
Considering these observations,
the DEM trends are consistent with a two-mechanism model
for the irreversible strains of the form
\begin{align}
  \label{eq:multi1}
  &d\mathbf{e}^{\text{(i)}} =
    d\mathbf{e}^{\text{(i)}}_{1} + d\mathbf{e}^{\text{(i)}}_{2} =
    \mathcal{F}^{\text{(i)}}_{1}(d\mathbf{s}) +
    \mathcal{F}^{\text{(i)}}_{2}(d\mathbf{s}) \\
  \label{eq:multi2}
  &\mathcal{F}^{\text{(i)}}_{a}(d\mathbf{s}) =
    [d\mathbf{e}_{a}^{\text{(i)}}] =
      \begin{cases}
        \displaystyle
        \frac{1}{h^{-}}[\mathbf{g}_{a}^{-}][\mathbf{f}_{a}]^{\mathsf{T}}[d\mathbf{s}], &
          [\mathbf{f}_{a}]^{\mathsf{T}}[d\mathbf{s}] \le 0 \\[1ex]
        \displaystyle
        \frac{1}{h^{+}}[\mathbf{g}_{a}^{+}][\mathbf{f}_{a}]^{\mathsf{T}}[d\mathbf{s}], &
          [\mathbf{f}_{a}]^{\mathsf{T}}[d\mathbf{s}] > 0
      \end{cases}
      \;,\quad a=1,2
\end{align}
with the subscript, $a=1$ or 2, denoting the particular mechanism.
Each mechanism has different plastic moduli,
$h^{+}$ and $h^{-}$, in the loading and unloading directions,
as suggested by generalized plasticity models
\cite{Rudnicki:1975a,Hashiguchi:2005a}.
Because of the axial symmetry of the initial triaxial loading,
the two unit yield directions, $\mathbf{f}_{1}$ and $\mathbf{f}_{2}$,
activate mechanisms
that are symmetric with respect to the
$d\vec{\mathbf{s}}_{1}$--$d\vec{\mathbf{s}}_{3}$ plane:
the two directions share the same orientation in
the $d\vec{\mathbf{s}}_{1}$--$d\vec{\mathbf{s}}_{3}$ plane
(such as the angle $\theta$ in Fig.~\ref{fig:circles}a) but have
mirrored directions that tilt out-of-plane
(i.e., the directions $\mathbf{f}_{1}$ and $\mathbf{f}_{2}$
are rotated by angles $\gamma$ and $-\gamma$
about the $d\vec{\mathbf{s}}_{1}$ axis,
Fig.~\ref{fig:circles}b).
Different unit flow directions, $\mathbf{g}^{+}$ and $\mathbf{g}^{-}$,
apply to loading and unloading, as is apparent from the spars
that emanate from the large and small circles
of Fig.~\ref{fig:circles}d.
The flow directions are in pairs, $(\mathbf{g}^{+}_{1},\mathbf{g}^{+}_{2})$
and $(\mathbf{g}^{-}_{1},\mathbf{g}^{-}_{2})$,
with each pair mirrored with respect to the
$d\vec{\mathbf{s}}_{1}$--$d\vec{\mathbf{s}}_{3}$ plane,
with orientation angles $\beta^{+}_{1}=\beta^{+}_{2}$
and $\beta^{-}_{1}=\beta^{-}_{2}$
in the $d\vec{\mathbf{s}}_{1}$--$d\vec{\mathbf{s}}_{2}$ plane
(Fig.~\ref{fig:circles}a), and tilted about the
$d\vec{\mathbf{s}}_{1}$ axis at the same angles $\gamma_{1}=-\gamma_{2}$
as are the yield directions (Fig.~\ref{fig:circles}b).
In the spaces of the generalized
stress and strain of Section~\ref{sec:generalized},
the yield direction
$\mathbf{f}_{a}=[\cos\theta_{a},\,\sin\theta_{a}\sin\gamma_{a},\, \sin\theta_{a}\cos\gamma_{a}]^{\mathsf{T}}$;
and the flow direction
$\mathbf{g}_{a}^{\pm}=[\cos\beta_{a}^{\pm},\,\sin\beta_{a}^{\pm}\sin\gamma_{a},\, \sin\beta_{a}^{\pm}\cos\gamma_{a}]^{\mathsf{T}}$.
\par
\begin{figure}
  \centering
  \includegraphics{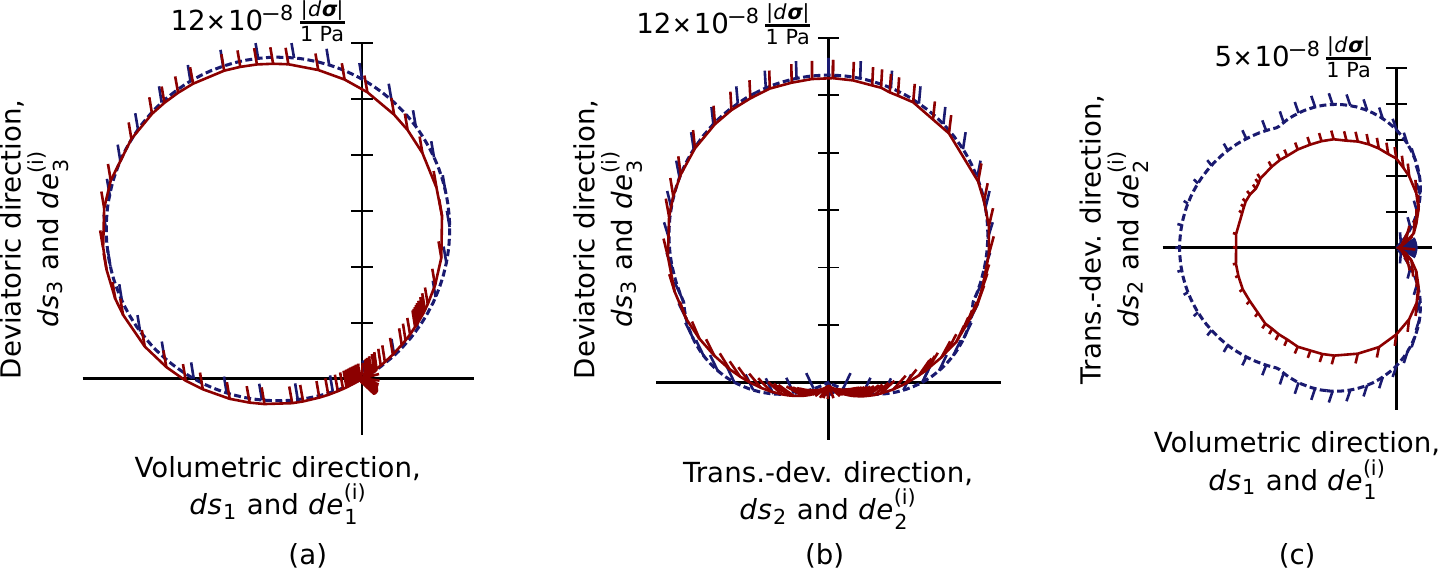}
  \caption{Comparison of DEM probes with the two-mechanism model
           of Eqs.~(\ref{eq:multi1}) and~(\ref{eq:multi2}),
           at strain $\varepsilon=-0.3\%$.
           Plots correspond to those in
           Figs.~\ref{fig:circles}a, \ref{fig:circles}b,
           and~\ref{fig:circles}c.
           Dashed lines (blue) are predictions of the two-mechanism model,
           with the best-fit parameters given in
           Table~\ref{table:multimechanism}.
           Solid red lines are the simulation result of
           Fig.~\ref{fig:circles}.
           The vertical rule and scale indicate the inelastic
           strain magnitude $|d\boldsymbol{\varepsilon}|$.
           \label{fig:multimechanism}}
\end{figure}

This generalized two-mechanism model
that includes irreversible unloading
reasonably fits the simulation data
(Fig.~\ref{fig:multimechanism}).
It's four tensorial zones are subtly apparent
as the circular lobes of the
dashed blue lines of Figs.~\ref{fig:multimechanism}b
and~\ref{fig:multimechanism}c.
The two mechanisms combine to create the features of
irreversible deformation that were found with the DEM simulations:
for example, at the strain $\varepsilon_{11}=-0.9\%$ of
Fig.~\ref{fig:circles},
a yield angle $\theta=121^{\circ}$
with respect to the isotropic axis,
a dilation direction $\beta=100^{\circ}$ with respect to this axis,
cusped corners,
a vanishing elastic region,
irreversible volumetric yielding, and the dependence of the flow direction
on the direction of the stress increment.
The best-fit parameters of the two-mechanism model are summarized in
Table~\ref{table:multimechanism} for loading probes at
the strains 0.3\%, 0.9\%, and 5.3\%.
\begin{table}
  \centering
  \caption{Best-fit parameters for the two-mechanism
           model of Eqs.(\ref{eq:multi1}) and~(\ref{eq:multi2})
           for the simulation results with Cattaneo--Mindlin contacts.
           Model results are shown in Fig.~\ref{fig:multimechanism}.
           \label{table:multimechanism}}
  \begin{tabular}{lcccccc}
    \toprule
      & \multicolumn{2}{c}{Hardening}
      & \multicolumn{2}{c}{Yield direction, $\mathbf{f}$}
      & \multicolumn{2}{c}{Flow direction, $\mathbf{g}$} \\
    \cmidrule(r){2-3} \cmidrule(r){4-5} \cmidrule(r){6-7}
    Strain & $h^{+}$ & $h^{-}$ & $\theta_{1}=\theta_{2}$ & $\gamma_{1}=-\gamma_{2}$
           & $\beta^{+}_{1}=\beta^{+}_{2}$
           & $\beta^{-}_{1}=\beta^{-}_{2}$\\
    \midrule
    0.3\% & 11.2 MPa & 1,650 MPa & 113$^{\circ}$ & 25$^{\circ}$ & \ \ 90.2$^{\circ}$ & $-$41$^{\circ}$\\
    0.9\% &  \ \ 6.6 MPa & 830 MPa & 119$^{\circ}$ & 26$^{\circ}$ & \ \ 99.4$^{\circ}$ & $-$41$^{\circ}$\\
    5.3\% &  \ \ 4.4 MPa & 360 MPa & 128$^{\circ}$ & 26$^{\circ}$ & 117.5$^{\circ}$ & $-$37$^{\circ}$\\
    \bottomrule
  \end{tabular}
\end{table}
The fit in Fig.~\ref{fig:multimechanism}
is certainly not perfect, with an average mismatch in the strains
$d\mathbf{e}^{\text{(i)}}$ that is 14\% of the average
magnitude $|d\mathbf{e}^{\text{(i)}}|$.
Figure~\ref{fig:multimechanism}c shows that this error
originates primarily in the model's overestimation of
the deviatoric strains $de_{3}^{\text{(i)}}$ when the deviatoric
stress $ds_{3}$ is close to zero.
The model also introduces subtle new corners that
were not observed in the simulations:
the slight corners seen in the second and third quadrants
of Fig.~\ref{fig:multimechanism}c.
\subsection{Tangential plasticity model}
The alternative tangential plasticity model
of Rudnicki and Rice \cite{Rudnicki:1975a}
has a primary yield direction, say $\mathbf{f}$,
but additional irreversible
deformation is induced by tangential
stress increments that are orthogonal
to this direction.
In the context of our results for an axisymmetric
initial loading, direction $\mathbf{f}$ lies in the
$\vec{\mathbf{s}}_{1}$--$\vec{\mathbf{s}}_{3}$ Rendulic plane
(for example, for the results at strain $\varepsilon_{11}=-0.9\%$
of Fig.~\ref{fig:circles}a, $\mathbf{f}$ is directed
at angle $\theta=121^{\circ}$),
and the tangential direction
is the transverse-deviatoric direction $\vec{\mathbf{s}}_{2}$.
A fairly general form for
the additional deformation produced by tangential loading would
be as follows:
the sum of a tangential flow increment that is symmetric in the unit
$\vec{\mathbf{s}}_{2}$ direction,
$(1/h_{1})\vec{\mathbf{s}}_{2}(\vec{\mathbf{s}}_{2} \cdot d\mathbf{s})$,
and a second flow increment that is anti-symmetric,
$(1/h_{2})\mathbf{g}_{2}|\vec{\mathbf{s}}_{2} \cdot d\mathbf{s}|$,
where unit flow direction $\mathbf{g}_{2}$ lies in
the $\vec{\mathbf{s}}_{1}$--$\vec{\mathbf{s}}_{3}$ plane.
Both of these irreversible flows are induced by loading in the
transverse $\vec{\mathbf{s}}_{2}$ direction
(i.e. from the projected stress increment
$\vec{\mathbf{s}}_{2} \cdot d\mathbf{s}$).
The anti-symmetric term in this model introduces an
additional yield direction besides the primary
direction $\mathbf{f}$, and as such, is a multi-mechanism model
that also includes a symmetric (and linear) flow term,
similar to the linear part of hypo-plastic models.
Like the two-mechanism model described above,
the tangential model includes additional tensorial zones
and introduces two additional parameters.
Despite its complexity, we found the fit to be less favorable
than the simple two-mechanism model, and
the average mismatch in the irreversible strains
$d\boldsymbol{\varepsilon}^{\text{(i)}}$ was 17\%
of the average strain magnitude $|d\boldsymbol{\varepsilon}^{\text{(i)}}|$.
%
\section{Conclusions}
The simulations reveal several fundamental
aspects of incremental granular behavior,
some of which deviate substantially
from conventional elasto-plasticity and have not been
reported in previous studies:
\begin{enumerate}[listparindent=2em]
\item
The reversible response that is produced by the
initial axisymmetric triaxial loading
conforms to an anisotropic linear
model with transverse isotropy, with an exception noted below.
The initial monotonic loading induced progressively
more anisotropic reversible (elastic) stiffness
with increased loading.
Although this result is not surprising,
the measured anisotropy was quite large,
with the axial Young's modulus becoming as much
as 6 times greater
than the modulus in lateral directions.
The anisotropy of the reversible stiffness, which
increases with increasing strain, is
consistent with the observed volumetric behavior of sands:
the reversible strain, by itself,
would transition from contractive to dilative
under constant lateral stress or would
transition from neutral to dilative under constant-$p$ loading.
\item
Although the reversible (elastic) stiffness operator is symmetric
at the start of loading, the incremental
stiffness soon becomes asymmetric,
with an asymmetry that grows as loading proceeds.
The elastic
compliance $\bar{C}_{ij}$ becomes greater than $\bar{C}_{ji}$
when ``$i$'' and ``$j$'' are the loading and transverse directions.
\item
Irreversible strain occurs for all directions of
the stress increment (loading, unloading, transverse-loading, etc.),
and, as has been stated,
a reversible domain, if it exists
at all, is smaller than the very small strain increments
of our simulations, $2\times 10^{-6}$.
In particular, a small irreversible strain even occurs
when the initial axisymmetric triaxial loading is reversed
(from triaxial compression to triaxial extension).
That is, irreversible strains occur for all directions of
stress increments within the $p$--$q$ Rendulic plane, abrogating
the conventional concept of opposite loading (yield) and
unloading (elastic) directions.
Moreover, the irreversible strain during a reversed increment
of triaxial extension is contractive, even as
the irreversible strain is dilative during triaxial compression.
\item
Following an initial loading along a path of axisymmetric triaxial
compression,
irreversible strains occur for stress increments
in the transverse-deviatoric direction,
and the deformation has isotropic, deviatoric, and
transverse-deviatoric components.
This transverse-deviator direction is tangent to the primary
yield surface that would develop during the initial axisymmetric
loading,
and the observed
irreversible strains are direct evidence of the tangential
plasticity that was theorized by
Rudnicki and Rice \cite{Rudnicki:1975a} and
Hashiguchi \cite{Hashiguchi:2005a}.
Large irreversible strain increments occur in this
transverse-deviatoric (tangential) direction for a wide
range of stress increment directions,
and these irreversible increments have both
volumetric and deviatoric components.
\item
In the three planes formed by the axes of
volumetric, deviatoric, and transverse-deviatoric stress,
the yield surface displays rounded corners.
Because the corners are rounded and not abrupt,
the irreversible strains change in a continuous manner
as the stress increment is varied near a corner.
\item
The direction of the irreversible strain increment
depends upon the
direction of the stress increment: a single
flow direction does not apply with granular materials.
For stress increments within the octahedral pi-plane,
the irreversible strain increments range from
contractive to dilative for different stress directions.
\item
During the initial monotonic constant-$p$ triaxial compression
in which the mean stress was kept constant, the
volume change in our simulations of a medium-dense
sand was initially contractive but transitioned
to dilative at larger strains
(the transition from contractive to dilative behavior,
referred to as the characteristic state,
occurred at a strain $\varepsilon_{11}=0.1\%$).
Under constant-$p$ conditions,
these volume tendencies are the combined result
of reversible and irreversible strain increments.
The intense anisotropy that is developed in the reversible stiffness
produces dilation throughout the loading process;
whereas, the irreversible response was contractive at small
strains but transitioned to dilative at larger strains.
The strong dilation at large strains is produced by both
reversible and irreversible responses.
%
%
\item
The irreversible strains that were measured
with DEM stress probes reasonably conform
to a two-mechanism elasto-plasticity model, although the fit
is not perfect.
The yield and flow directions of each mechanism must
have a component in the transverse-deviatoric
direction that can produce the tangential irreversible strains
that were measured with the simulations.
The model must also include irreversible deformations in
both loading and unloading directions.
Our attempt to fit a more complex tangential plasticity model
was less successful, due to
the complex coupling of the three components of the irreversible
strains.
\end{enumerate}
Taken together, the results contradict five of the six
principles of conventional elasto-plasticity that were
given in the Introduction,
and the results go further,
by also revealing details of this contravening behavior.
\par
We note that
these results were derived from simulations with an assembly
having Cattaneo--Mindlin contacts, which permit elastic,
micro-slip, and slip behaviors at the contact level.
The authors conducted similar simulations with conventional
linear-frictional contacts,
and all aspects of the results enumerated
above were qualitatively the same as those
with Cattaneo--Mindlin contacts.
\par
Our study was limited to behavior that followed
an initial monotonic
drained triaxial compression, and
this work could certainly be extended to more general loading paths:
other proportional loading paths, paths involving principal stress
rotations, and cyclic paths.
The results suggest, however, that granular behavior is more complex
than previously thought and that granular materials should be
treated as thoroughly inelastic and
anisotropic materials
that can only be modeled, even approximately,
with more advanced constitutive frameworks
than conventional elasto-plasticity.
%
\pagebreak
\appendix
\section{DEM modeling details and verification}\label{sec:DEM}
\begin{figure}
  \centering
  \includegraphics[scale=0.20]{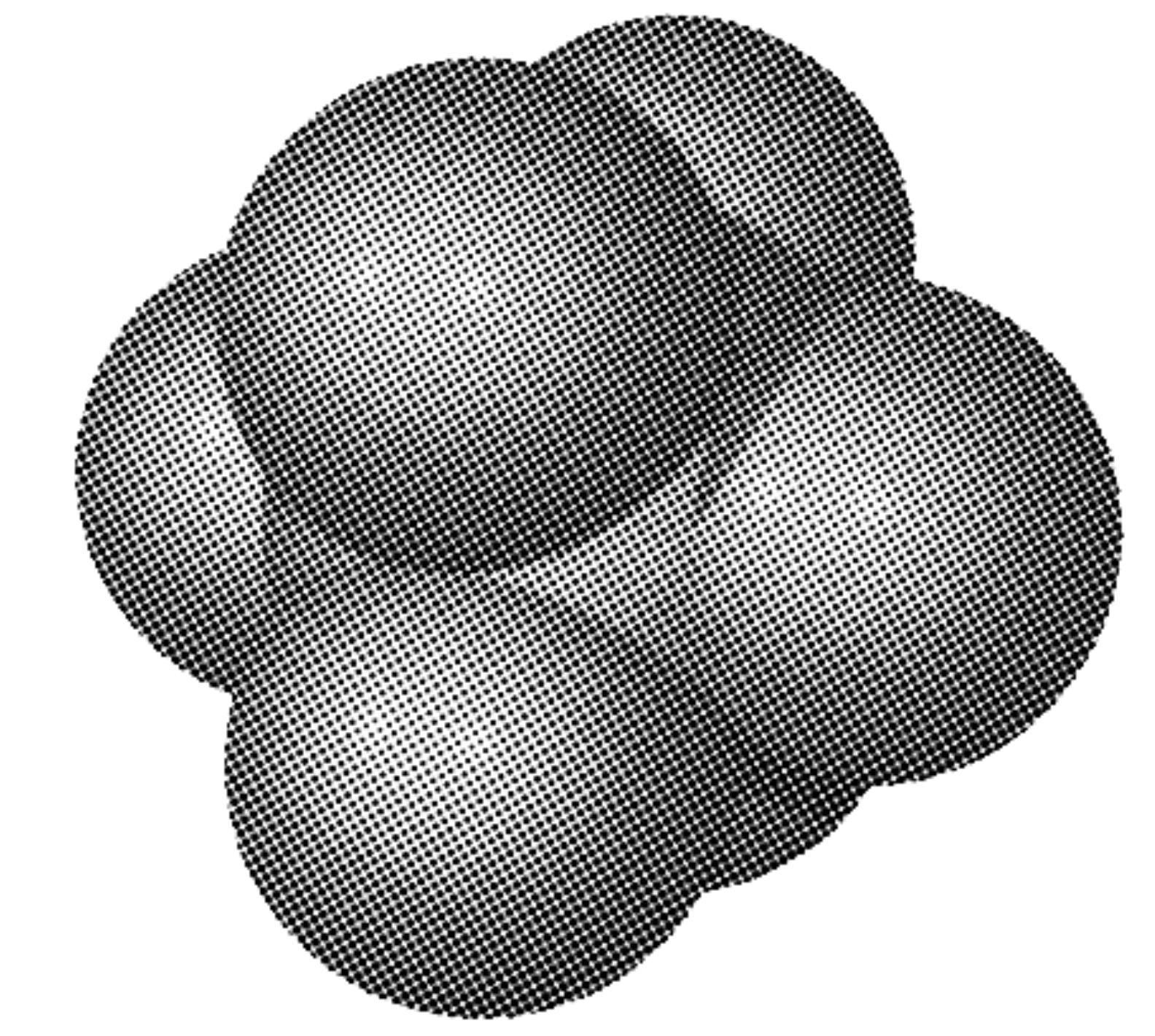}
  \caption{Sphere-cluster particle.
           \label{fig:spherecluster}}
\end{figure}

Discrete element
simulations were conducted with a cubical assembly of 10,648 particles contained
within periodic boundaries.
The primary DEM algorithm is that of Cundall and Strack \cite{Cundall:1979a},
in which servo-controlled boundaries were used to maintain
a constant mean stress during the initial monotonic loading
and selective combinations of the principal stresses during
stress probes \cite{Kuhn:1992a,Thornton:2000a}.
The simulations
were intended to produce a modest fidelity to the
bulk behavior of sands at low mean stress.
Because sphere assemblies produce unrealistic rolling between
particles and have a low bulk strength, we used a bumpy,
non-convex cluster shape for the particles:
a large central sphere with six smaller embedded outer
spheres in an octahedral arrangement (Fig.~\ref{fig:spherecluster}).
The use of non-spherical particles circumvents the need
of artificial measures to restrain the particle rotation
\cite{Vuquoc:2000a,Favier:2001a}
(circumventing, for example, the use of rotational contact springs or
the direct restraint of particle rotations,
as in \cite{Plassiard:2009a}).
Through trial and error, we chose the radii of the central and outer
spheres so that the bulk behavior approximated that of Nevada
Sand, a standard poorly graded sand (SP) use in laboratory
and centrifuge testing programs
\cite{Arulmoli1992a,Kuhn:2014c}.
The particle size range was 0.074--0.28~mm,
with $D_{50}=0.165$~mm,
with a size distribution that closely matched that of Nevada Sand.
\par
To create assemblies with a range of densities,
we began with the particles sparsely and randomly
arranged within a cubic periodic cell.
With an initial low interparticle friction coefficient
($\mu = 0.20$), the assembly was isotropically reduced in
dimension until it ``seized'' when a sufficiently complete
contact network had formed.
A series of 25 progressively denser assemblies were created
by repeatedly assigning random velocities to the particles
of the previous assembly and then further reducing the cell's
dimensions until it, too, had seized.
The 25 assemblies had void ratios ranging from
0.781 to 0.586 (solid fractions of 0.561 to 0.664).
The single assembly used in the paper
had void ratio 0.690
(solid fraction 0.592)
and approximates the behavior of Nevada Sand at a relative
density of 40\%.
After compaction, the assembly was allowed to quiesce with
friction coefficient $\mu=0.40$, which was then raised
to $\mu=0.55$ for the subsequent loading simulations
(a value that produced drained and undrained behaviors
similar to Nevada Sand \cite{Kuhn:2014c}).
\par
The particles are durable (non-breaking) and interact
only at their contacts.
Normal forces were modeled with a generalization of Hertz theory
\cite{Thornton:1988a},
and tangential forces were modeled with a full implementation
of the complementary Cattaneo--Mindlin contact.
The modified Hertz contact
has the advantage of producing a small-strain shear
modulus that is pressure-dependent,
with a small-strain shear modulus that is proportional
to pressure $p$ raised to power 0.5, as is commonly observed
in laboratory and field testing of sands \cite{Kuhn:2014c}.
With regard to tangential contact forces,
approximate implementations
of the Hertz contact typically
use a tangential stiffness that is proportional to the normal
force and not upon the directions of the normal
and tangential movements \cite{Cundall:1988a,Lin:1997a}.
Such
implementations do not permit frictional dissipation in the form
of micro-slip and can
result in an unfortunate and physically inadmissible infusion
of energy during close cycles of
contact movement \cite{Elata:1996a}.
To avoid these problems,
we used the
J\"{a}ger algorithm for the tangential
force, which can model arbitrary sequences
of normal and tangential contact movements \cite{Kuhn:2011a}.
With the Cattaneo--Mindlin contacts,
the loading simulations were conducted with
an inter-particle friction coefficient
$\mu=0.55$,
particle shear modulus $G=29$~GPa, and Poisson ratio $\nu=0.15$.
We used a scaling of of the particles' material density,
in the manner of Thornton \cite{Thornton:2000a}, so that slow,
quasi-static simulations could be completed with reasonable
computation time.
\par
As has been noted by Alonso-Marroqu\'{i}n et al. \cite{AlonsoMarroquin:2005b}
and Froiio and Roux \cite{Froiio:2010a},
DEM simulations necessarily involve a compromise between scientific
intent and computational expedience.
Without proper care, DEM simulations can yield results
that are sensitive to strain rate, due to the particles' inertias
and to the
damping that is employed to stabilize
the particles' motions \cite{Suzuki:2014a}.
Our intent was to model behavior in which these effects
were minimized so that rate-independent behavior was attained.
To this end, we used a slow strain rate
(strain increments of $1\times 10^{-8}$
and minimal viscous damping (2\% of critical damping).\par
Several performance parameters were used to verify
the quasi-static and rate-independent nature of these probes.
The inertial number $I=\dot{\varepsilon}\sqrt{m/(pd)}$,
a relative measure of loading and inertial rates,
was about $1\times 10^{-11}$,
signifying nearly quasi-static loading
\cite{daCruz:2005a}.
During the incremental probes,
the average imbalance of force on a particle was less than
0.003\% of the average contact force
(parameter $\chi$ in \cite{Ng:2006a,Suzuki:2014a}).
The average kinetic energy of the particles was less than
$3\times 10^{-7}$ times the
elastic energy in the contacts.
With the very slow strain rates of the simulations,
doubling the strain increment from $1\times10^{-8}$
to $2\times10^{-8}$ had minimal effect on the
monotonic stress-strain response. 
Boundary movements were regulated so that any six of the
stress or strain components (or any six linear combination of these
components) could be controlled at desired rates.
When a stress component was controlled, it would typically remain
on target to within 0.001~Pa,
compared with the mean stress of 100~kPa and stress-probes
that produced stress changes $|d\sigma_{ij}|$ on the order
of 100~Pa.
As further verification of strain rate indifference during
loading,
we also conducted brief creep and stress relaxation tests in
which either the stress or the assembly boundaries were frozen
at the end of a stress probe.
Froiio and Roux \cite{Froiio:2010a} have noted a tendency
for an assembly to exhibit creep during small stress probes,
an inclination that can obscure the probe results.
During our creep tests, the strain rate was
0.5--3$\times 10^{-10}$, far less than the loading rate of $1\times10^{-8}$.
During stress relaxation with zero strain rate, the stress changed at
a rate of less than 4\% of that typically measured during
the incremental stress probes that were used in this
study.
The strain increment of $2\times 10^{-6}$ and strain steps of
$1\times 10^{-8}$ entailed 200 steps.
Although small fluctuations were noted in the first few steps,
the advance of stress and strain were fairly uniform within
the 200 strain steps.
Moreover,
the envelopes in
Figs.~\ref{fig:piplane}--\ref{fig:s1s2}
and~\ref{fig:cusps} are quite smooth,
even near the origin of these plots,
where the strain parts are minuscule, and at high magnification,
indicating that random errors are negligible.
All of these measurements indicate that the behavior in the
simulations was nearly quasi-static and independent of loading rate.
\par
The DEM simulations were done with the authors'
OVAL code (see \cite{Kuhn:2014c}) and
were run with a 4th generation Intel i7 processor
on a single thread.
The monotonic loading in Fig.~\ref{fig:StressStrain},
in which the strain was advanced to 16\%, took
47~days of compute time; whereas, each probe
in the series of Figs.~\ref{fig:piplane} to~\ref{fig:s1s2}
took about 6~minutes of compute time.
\par
\pagebreak

\begin{thebibliography}{10}
\expandafter\ifx\csname url\endcsname\relax
  \def\url#1{\texttt{#1}}\fi
\expandafter\ifx\csname urlprefix\endcsname\relax\def\urlprefix{URL }\fi
\expandafter\ifx\csname href\endcsname\relax
  \def\href#1#2{#2} \def\path#1{#1}\fi

\bibitem{Yang:2005a}
Y.~Yang, H.~S. Yu, K.~K. Muraleetharan, Solution existence conditions for
  elastoplastic constitutive models of granular materials, Int. J. Plasticity
  21~(12) (2005) 2406--2425.

\bibitem{Wu:1996a}
W.~Wu, E.~Bauer, D.~Kolymbas, Hypoplastic constitutive model with critical
  state for granular materials, Mech. of Mater. 23~(1) (1996) 45--69.

\bibitem{Lin:2015a}
J.~Lin, W.~Wu, R.~I. Borja, Micropolar hypoplasticity for persistent shear band
  in heterogeneous granular materials, Comput. Methods Appl. Mech. Eng. 289
  (2015) 24--43.

\bibitem{Hashiguchi:2005a}
K.~Hashiguchi, Generalized plastic flow rule, Int. J. Plasticity 21~(2) (2005)
  321--351.

\bibitem{Zhu:2010a}
Q.~Z. Zhu, J.-F. Shao, M.~Mainguy, A micromechanics-based elastoplastic damage
  model for granular materials at low confining pressure, Int. J. Plasticity
  26~(4) (2010) 586--602.

\bibitem{Nicot:2011c}
F.~Nicot, F.~Darve, The {H}-microdirectional model: accounting for a mesoscopic
  scale, Mech. of Mater. 43~(12) (2011) 918--929.

\bibitem{Yeh:2006a}
W.-C. Yeh, H.-Y. Lin, An endochronic model of yield surface accounting for
  deformation induced anisotropy, Int. J. Plasticity 22~(1) (2006) 16--38.

\bibitem{Ganzenmuller:2011a}
G.~C. Ganzenm{\"u}ller, S.~Hiermaier, M.~O. Steinhauser, Shock-wave induced
  damage in lipid bilayers: a dissipative particle dynamics simulation study,
  Soft Matter 7~(9) (2011) 4307--4317.

\bibitem{Lewin:1970a}
P.~I. Lewin, J.~B. Burland, Stress-probe experiments on saturated normally
  consolidated clay, G{\'{e}}otechnique 20~(1) (1970) 38--56.

\bibitem{Tatsuoka:1974b}
F.~Tatsuoka, K.~Ishihara, Yielding of sand in triaxial compression, Soils and
  Found. 14~(2) (1974) 63--76.

\bibitem{Bardet:1994a}
J.~P. Bardet, Observations on the effects of particle rotations on the failure
  of idealized granular materials, Mech. of Mater. 18~(2) (1994) 159--182.

\bibitem{Anandarajah:1995a}
A.~Anandarajah, K.~Sobhan, N.~Kuganenthira, Incremental stress-strain behavior
  of granular soil, J. Geotech. Eng. 121~(1) (1995) 57--68.

\bibitem{Royis:1998a}
P.~Royis, T.~Doanh, Theoretical analysis of strain response envelopes using
  incrementally non-linear constitutive equations, Int. J. Numer. and Anal.
  Methods in Geomech. 22~(2) (1998) 97--132.
\newblock \href
  {http://dx.doi.org/10.1002/(SICI)1096-9853(199802)22:2<97::AID-NAG911>3.0.CO;2-Z}
  {\path{doi:10.1002/(SICI)1096-9853(199802)22:2<97::AID-NAG911>3.0.CO;2-Z}}.

\bibitem{Kishino:2003a}
Y.~Kishino, On the incremental nonlinearity observed in a numerical model for
  granular media, Revista Italiana Di Geotecnica 3 (2003) 30--38.

\bibitem{AlonsoMarroquin:2005a}
F.~Alonso-Marroqu{\'{i}}n, H.~Herrmann, The incremental response of soils. an
  investigation using the discrete-element method, J. Eng. Math. 52~(1-3)
  (2005) 11--34.

\bibitem{AlonsoMarroquin:2005b}
F.~Alonso-Marroqu{\'{i}}n, S.~Luding, H.~J. Herrmann, I.~Vardoulakis, Role of
  anisotropy in the elastoplastic response of a polygonal packing, Phys. Rev. E
  71~(5) (2005) 051304.

\bibitem{Calvetti:2003a}
F.~Calvetti, G.~Viggiani, C.~Tamagnini, A numerical investigation of the
  incremental behavior of granualr soils, Rivista Italiana di Geotecnica 3
  (2003) 11--29.

\bibitem{Tamagnini:2005a}
C.~Tamagnini, F.~Calvetti, G.~Viggiani, An assessment of plasticity theories
  for modeling the incrementally nonlinear behavior of granular soils, Journal
  of engineering mathematics 52~(1-3) (2005) 265--291.

\bibitem{Sibille:2007a}
L.~Sibille, F.~Nicot, F.~V. Donz{\'e}, F.~Darve, Material instability in
  granular assemblies from fundamentally different models, Int. J. Numer. and
  Anal. Methods in Geomech. 31~(3) (2007) 457--481.

\bibitem{Plassiard:2009a}
J.-P. Plassiard, N.~Belheine, F.-V. Donz{\'e}, A spherical discrete element
  model: calibration procedure and incremental response, Granul. Matter 11~(5)
  (2009) 293--306.

\bibitem{Froiio:2010a}
F.~Froiio, J.-N. Roux, Incremental response of a model granular material by
  stress probing with dem simulations, in: J.~Goddard, P.~Giovine, J.~T.
  Jenkins (Eds.), IUTAM-ISIMM Symposium on mathematical modeling and physical
  instances of granular flow. AIP Conference Proceedings, Vol. 1227, AIP, 2010,
  pp. 183--197.

\bibitem{Harthong:2013a}
B.~Harthong, R.~G. Wan, Directional plastic flow and fabric dependencies in
  granular materials, in: AIP Conference Proceedings, Vol. 1542, 2013, pp.
  193--196.
\newblock \href {http://dx.doi.org/10.1063/1.4811900}
  {\path{doi:10.1063/1.4811900}}.

\bibitem{Wan:2014a}
R.~Wan, M.~Pinheiro, On the validity of the flow rule postulate for
  geomaterials, Int. J. Numer. and Anal. Methods in Geomech. 38~(8) (2014)
  863--880.

\bibitem{Kuhn:2018a}
M.~R. Kuhn, A.~Daouadji, Quasi-static incremental behavior of granular
  materials: Elastic–plastic coupling and micro-scale dissipation, J. Mech.
  Phys. Solids 114 (2018) 219--237.
\newblock \href {http://dx.doi.org/https://doi.org/10.1016/j.jmps.2018.02.019}
  {\path{doi:https://doi.org/10.1016/j.jmps.2018.02.019}}.

\bibitem{Agnolin:2007c}
I.~Agnolin, J.-N. Roux, Internal states of model isotropic granular packings.
  {III}. elastic properties, Phys. Rev. E 76 (2007) 061304.

\bibitem{Hueckel:1976a}
T.~Hueckel, Coupling of elastic and plastic deformations of bulk solids,
  Meccanica 11~(4) (1976) 227--235.
\newblock \href {http://dx.doi.org/10.1007/BF02128296}
  {\path{doi:10.1007/BF02128296}}.

\bibitem{Dafalias:1977b}
Y.~F. Dafalias, Elasto-plastic coupling within a thermodynamic strain space
  formulation of plasticity, Int. J. Non-linear Mech. 12~(5) (1977) 327--337.

\bibitem{Hueckel:1977a}
T.~Hueckel, G.~Maier, Incremental boundary value problems in the presence of
  coupling of elastic and plastic deformations: a rock mechanics oriented
  theory, Int. J. Solids Struct. 13~(1) (1977) 1--15.

\bibitem{Collins:1997a}
I.~F. Collins, G.~T. Houlsby, Application of thermomechanical principles to the
  modelling of geotechnical materials, Proc. R. Soc. Lond. A 453~(1964) (1997)
  1975--2001.
\newblock \href
  {http://arxiv.org/abs/http://rspa.royalsocietypublishing.org/content/453/1964/1975.full.pdf}
  {\path{arXiv:http://rspa.royalsocietypublishing.org/content/453/1964/1975.full.pdf}}.

\bibitem{Pastor:1986a}
M.~Pastor, O.~C. Zienkiewicz, A generalized plasticity, hierarchical model for
  sand under monotonic and cyclic loading, in: Proc., 2nd Int. Symp. on
  Numerical Models in Geomechanics, Jackson and Son, Ghent, Belgium, 1986, pp.
  131--150.

\bibitem{Hill:1967a}
R.~Hill, The essential structure of constitutive laws for metal composites and
  polycrystals, J. Mech. Phys. Solids 15~(2) (1967) 79--95.

\bibitem{Khojastehpour:2006a}
M.~Khojastehpour, Y.~Murakami, K.~Hashiguchi, Antisymmetric bifurcation in an
  elastoplastic cylinder with tangential plasticity, Mech. of Mater. 38~(11)
  (2006) 1061--1071.

\bibitem{Kuhn:2002b}
M.~R. Kuhn, {OVAL} and {OVALPLOT}: programs for analyzing dense particle
  assemblies with the {D}iscrete {E}lement {M}ethod,
  http://faculty.up.edu/kuhn/oval/oval.html.

\bibitem{Steinhauser:2009a}
M.~O. Steinhauser, K.~Grass, E.~Strassburger, A.~Blumen, Impact failure of
  granular materials---non-equilibrium multiscale simulations and high-speed
  experiments, Int. J. Plasticity 25~(1) (2009) 161--182.

\bibitem{Chakraborty:2013a}
S.~Chakraborty, A.~Shaw, A pseudo-spring based fracture model for sph
  simulation of impact dynamics, Int. J. Impact Engrg. 58 (2013) 84--95.

\bibitem{Kuhn:2014c}
M.~R. Kuhn, H.~Renken, A.~Mixsell, S.~Kramer, Investigation of cyclic
  liquefaction with discrete element simulations, J. Geotech. and Geoenv. Eng.
  140~(12) (2014) 04014075.
\newblock \href {http://dx.doi.org/10.1061/(ASCE)GT.1943-5606.0001181}
  {\path{doi:10.1061/(ASCE)GT.1943-5606.0001181}}.

\bibitem{Kuhn:2009b}
M.~R. Kuhn, K.~Bagi, Specimen size effect in discrete element simulations of
  granular assemblies, J. Eng. Mech. 135~(6) (2009) 485--492.

\bibitem{Mindlin:1953a}
R.~Mindlin, H.~Deresiewicz, Elastic spheres in contact under varying oblique
  forces, J. Appl. Mech. 19~(1) (1953) 327--344.

\bibitem{Kuhn:2011a}
M.~R. Kuhn, Implementation of the {J}{\"{a}}ger contact model for discrete
  element simulations, Int. J. Numer. Methods Eng. 88~(1) (2011) 66--82.

\bibitem{Elata:1996a}
D.~Elata, J.~G. Berryman, Contact force-displacement laws and the mechanical
  behavior of random packs of identical spheres, Mech. of Mater. 24~(3) (1996)
  229--240.

\bibitem{Nicot:2006a}
F.~Nicot, F.~Darve, \href{http://dx.doi.org/10.1007/s10035-006-0012-4}{On the
  elastic and plastic strain decomposition in granular materials}, Granul.
  Matter 8~(3-4) (2006) 221--237.
\newblock \href {http://dx.doi.org/10.1007/s10035-006-0012-4}
  {\path{doi:10.1007/s10035-006-0012-4}}.
\newline\urlprefix\url{http://dx.doi.org/10.1007/s10035-006-0012-4}

\bibitem{Hueckel:1979a}
T.~Hueckel, R.~Nova, Some hysteresis effects of the behaviour of geologic
  media, Int. J. Solids Struct. 15~(8) (1979) 625--642.

\bibitem{Pouragha:2017a}
M.~Pouragha, R.~Wan, Non-dissipative structural evolutions in granular
  materials within the small strain range, Int. J. Solids Struct. 110--111
  (2017) 94--105.

\bibitem{Pouragha:2018a}
M.~Pouragha, R.~Wan, On elastic deformations and decomposition of strain in
  granular media, Int. J. Solids Struct.

\bibitem{Rendulic:1936a}
L.~Rendulic, Relation between void ratio and effective principal stresses for a
  remolded silty clay, in: Proc. of 1st Int. Conf. on Soil Mechanics and
  Foundation Engineering, Vol.~3, Harvard University Press, Cambridge, MA,
  1936, pp. 48--51.

\bibitem{Gudehus:1979a}
G.~Gudehus, D.~Kolymbas, A constitutive law of rate type for soils, in:
  W.~Wittke (Ed.), Numerical Methods in Geomechanics Aachen 1979, Vol.~1, A. A.
  Balkema, Rotterdam, 1979, pp. 319--329.

\bibitem{Darve:2005b}
F.~Darve, F.~Nicot, On flow rule in granular media: phenomenological and
  multi-scale views ({P}art {II}), Int. J. Numer. and Anal. Methods in Geomech.
  29~(14) (2005) 1411--1432.

\bibitem{Hill:1959a}
R.~Hill, Some basic principles in the mechanics of solids without a natural
  time, J. Mech. Phys. Solids 7~(3) (1959) 209--225.

\bibitem{Darve:1990a}
F.~Darve, The expression of rheological laws in incremental form and the main
  classes of constitutive equations, in: F.~Darve (Ed.), Geomaterials:
  constitutive equations and modelling, Elsevier, London, 1990, pp. 123--147.

\bibitem{Zienkiewicz:1984a}
O.~C. Zienkiewicz, Z.~Mroz, Generalized plasticity formulation and applications
  to geomechanics, in: C.~S. Desai, R.~H. Gallagher (Eds.), Mechanics of
  engineering materials, Vol.~44, Wiley, New York, 1984, pp. 655--680.

\bibitem{Hashiguchi:1993a}
K.~Hashiguchi, Fundamental requirements and formulation of elastoplastic
  constitutive equations with tangential plasticity, Int. J. Plasticity 9~(5)
  (1993) 525--549.

\bibitem{Lai:2016a}
Y.~Lai, M.~Liao, K.~Hu, A constitutive model of frozen saline sandy soil based
  on energy dissipation theory, Int. J. Plasticity 78 (2016) 84--113.

\bibitem{Rudnicki:1975a}
J.~W. Rudnicki, J.~R. Rice, Conditions for the localization of deformation in
  pressure-sensitive dilatant materials, J. Mech. Phys. Solids 23 (1975)
  371--394.

\bibitem{Darve:2005a}
F.~Darve, F.~Nicot, On incremental non-linearity in granular media:
  phenomenological and multi-scale views (part i), Int. J. Numer. and Anal.
  Methods in Geomech. 29~(14) (2005) 1387--1409.

\bibitem{Hicher:2007a}
P.-Y. Hicher, C.~S. Chang, A microstructural elastoplastic model for
  unsaturated granular materials, Int. J. Solids Struct. 44~(7-8) (2007)
  2304--2323.

\bibitem{Iai:2011a}
S.~Iai, T.~Tobita, O.~Ozutsumi, K.~Ueda, Dilatancy of granular materials in a
  strain space multiple mechanism model, Int. J. Numer. and Anal. Methods in
  Geomech. 35~(3) (2011) 360--392.

\bibitem{Tordesillas:2008a}
A.~Tordesillas, M.~Muthuswamy, S.~D.~C. Walsh, Mesoscale measures of nonaffine
  deformation in dense granular assemblies, J. Eng. Mech. 134~(12) (2008)
  1095--1113.

\bibitem{Kuhn:2016a}
M.~R. Kuhn, Maximum disorder model for dense steady-state flow of granular
  materials, Mech. of Mater. 93 (2016) 63--80.

\bibitem{Kuhn:2016b}
M.~R. Kuhn, Contact transience during slow loading of dense granular materials,
  J. Eng. Mech. (2016) C4015003\href
  {http://dx.doi.org/10.1061/(ASCE)EM.1943-7889.0000992}
  {\path{doi:10.1061/(ASCE)EM.1943-7889.0000992}}.

\bibitem{Pietruszczak:1986a}
S.~Pietruszczak, A flow theory for soil: Concept of multiple neutral loading
  surfaces, Comput. and Geotech. 2~(3) (1986) 185--203.

\bibitem{Anandarajah:2008a}
A.~Anandarajah, Multi-mechanism anisotropic model for granular materials, Int.
  J. Plasticity 24~(5) (2008) 804--846.

\bibitem{Mandel:1965a}
J.~Mandel, G{\'e}n{\'e}ralisation de la th{\'e}orie de plasticit{\'e} de {WT}
  {K}oiter, Int. J. Solids Struct. 1~(3) (1965) 273--295.

\bibitem{Cundall:1979a}
P.~A. Cundall, O.~D.~L. Strack, A discrete numerical model for granular
  assemblies, G{\'{e}}otechnique 29~(1) (1979) 47--65.

\bibitem{Kuhn:1992a}
M.~R. Kuhn, J.~K. Mitchell, Modelling of soil creep with the discrete element
  method, Eng. Comput. 9~(2) (1992) 277--287.

\bibitem{Thornton:2000a}
C.~Thornton, Numerical simulations of deviatoric shear deformation of granular
  media, G{\'{e}}otechnique 50~(1) (2000) 43--53.

\bibitem{Vuquoc:2000a}
L.~Vu-Quoc, X.~Zhang, O.~R. Walton, A 3-{D} discrete-element method for dry
  granular flows of ellipsoidal particles, Comput. Methods Appl. Mech. Eng.
  187~(3) (2000) 483--528.

\bibitem{Favier:2001a}
J.~F. Favier, M.~H. Abbaspour-Fard, M.~Kremmer, Modeling nonspherical particles
  using multisphere discrete elements, J. Eng. Mech. 127~(10) (2001) 971--977.

\bibitem{Arulmoli1992a}
K.~Arulmoli, K.~K. Muraleetharan, M.~M. Hossain, L.~S. Fruth, {VELACS}
  verification of liquefaction analyses by centrifuge studies laboratory
  testing program soil data report, Tech. Rep. Project No. 90-0562, The Earth
  Technology Corporation, Irvine, CA, data available through
  http://yees.usc.edu/velacs (1992).

\bibitem{Thornton:1988a}
C.~Thornton, C.~W. Randall, Applications of theoretical contact mechanics to
  solid particle system simulation, in: M.~Satake, J.~Jenkins (Eds.),
  Micromechanics of Granular Materials, Elsevier Science Pub. B.V., Amsterdam,
  The Netherlands, 1988, pp. 133--142.

\bibitem{Cundall:1988a}
P.~A. Cundall, Computer simulations of dense sphere assemblies, in: M.~Satake,
  J.~Jenkins (Eds.), Micromechanics of Granular Materials, Elsevier Science
  Pub. B.V., Amsterdam, The Netherlands, 1988, pp. 113--123.

\bibitem{Lin:1997a}
X.~Lin, T.-T. Ng, A three-dimensional discrete element model using arrays of
  ellipsoids, G{\'{e}}otechnique 47~(2) (1997) 319--329.

\bibitem{Suzuki:2014a}
K.~Suzuki, M.~R. Kuhn, Uniqueness of discrete element simulations in monotonic
  biaxial shear tests, Int. J. Geomech. 14~(5) (2014) 06014010.
\newblock \href {http://dx.doi.org/10.1061/(ASCE)GM.1943-5622.0000365}
  {\path{doi:10.1061/(ASCE)GM.1943-5622.0000365}}.

\bibitem{daCruz:2005a}
F.~da~Cruz, S.~Emam, M.~Prochnow, J.-N. Roux, F.~Chevoir, Rheophysics of dense
  granular materials: Discrete simulation of plane shear flows, Phys. Rev. E
  72~(2) (2005) 021309.

\bibitem{Ng:2006a}
T.-T. Ng, Input parameters of discrete element methods, J. Eng. Mech. 132~(7)
  (2006) 723--729.

\end{thebibliography}
\bibliographystyle{elsarticle-num}

\end{document}